\documentclass[%
reprint,
superscriptaddress,
showkeys,
amsmath,amssymb,
aps,
]{revtex4-2}

\usepackage{graphicx}
\usepackage{dcolumn}
\usepackage{bm}

\usepackage{siunitx}
\usepackage{color}
\usepackage[dvipsnames]{xcolor}

\newcommand{\avg}[1]{\left\langle #1\right\rangle}

\DeclareSIUnit{\einstein}{E}

\begin{document}
\title{Single-cell metabolic flux analysis reveals coexisting optimal sub-groups, cross-feeding, and mixotrophy in a cyanobacterial population}


\author{Arián Ferrero-Fernández}
\thanks{These authors contributed equally to this work.}
\affiliation{Biofisika Institute (CSIC, UPV/EHU), Barrio Sarriena s/n, 48940 Leioa, Bizkaia, Spain}  
 \affiliation{Department of Molecular Biology and Biochemistry, University of the Basque Country, Leioa, Bizkaia,
Spain. }

\author{Paula Prondzinsky}
\thanks{These authors contributed equally to this work.}
\affiliation{
 Japan Agency for Marine-Earth Science and Technology, Yokosuka 237-0061, Japan
}%

\affiliation{
 Earth-Life Science Institute, Institute of Science Tokyo, Tokyo 152-8550, Japan}%

\author{Lucia Gastoldi}
\affiliation{Water Research Center, New York University Abu Dhabi, Abu Dhabi, United Arab Emirates}

\author{David A. Fike}
\affiliation{Earth, Environmental, and Planetary Sciences, Washington University in St. Louis, St. Louis, MO 63130 USA}

\author{Harrison B. Smith}
\affiliation{
 Earth-Life Science Institute, Institute of Science Tokyo, Tokyo 152-8550, Japan}
\affiliation{
 Blue Marble Space Institute of Science, Seattle, WA, USA
}%

\author{Daniele De Martino}%
 \email{daniele.demartino@ehu.eus}
  \affiliation{Biofisika Institute (CSIC, UPV/EHU), Barrio Sarriena s/n, 48940 Leioa, Bizkaia, Spain}  
\affiliation{Ikerbasque Foundation, Bilbao, Spain
}


\author{Andrea De Martino}
\email{andrea.demartino@polito.it} 
\affiliation{
 Politecnico di Torino, Corso Duca degli Abruzzi 24, 10129 Torino, Italy
}%
\affiliation{
 Italian Insitute for Genomic Medicine, SP142 95, 10060 Candiolo, Italy }%

\author{Shawn Erin McGlynn}
 \email{mcglynn@elsi.jp}
\affiliation{
 Earth-Life Science Institute, Institute of Science Tokyo, Tokyo 152-8550, Japan}
\affiliation{
 Blue Marble Space Institute of Science, Seattle, WA, USA
}%



\begin{abstract}
We derive a single-cell level understanding of metabolism in an isogenic cyanobacterial population by integrating secondary ion mass spectrometry (SIMS) derived multi-isotope uptake measurements of \textit{Synechocystis} sp. PCC6803 with a statistical inference protocol based on Liebig’s law of the minimum, the maximum entropy principle, and constraint-based modeling. We find the population is structured in two metabolically distinct clusters: cells optimizing carbon yield while excessively turning over nitrogen, and cells which act reciprocally, optimizing nitrogen yield and excessively turning over carbon. This partition enables partial heterotrophy within the population via metabolic exchange, likely in the form of organic acids. Exchange increases the feasible metabolic space, and mixotrophic cells achieve the fastest growth rates. Metabolic flux analysis at the single-cell level reveals heterogeneity in carbon fixation rates, Rubisco specificity, and nitrogen assimilation. Our results provide a necessary foundation for understanding how population level phenotypes arise from the collective contributions of distinct individuals.  
\end{abstract}

\keywords{Cyanobacteria, metabolic flux analysis, cross-feeding, exudate, statistical inference, networks, metabolite exchange, single-cell heterogeneity}


\maketitle


\section*{Introduction}


The coexistence of identical states is statistically improbable. In line with this, genetically identical cells grown in the same environment can exhibit striking differences in phenotypes \cite{Ackermann2013, Ackermann2015, Schreiber2016, Zimmermann2018, Sheik2016, Hermelink2009, Berthelot2019, Zimmermann2015, Calabrese2019}. Advances in single-cell RNA sequencing have provided a dramatic view into this \cite{scell1,scell2,scell3,rosenthal2018}, contributing to the emerging picture that populations are composed of subpopulations that are phenotypically distinct. However, transcript levels alone do not translate directly into metabolic activity due to the complex relationship between gene expression, enzyme abundance and reaction fluxes \cite{ishii2007multiple, taniguchi2010quantifying}. Achieving a more direct understanding of metabolic phenotypes will be crucial to develop a more complete ecological understanding of microbial populations, as well as for our ability to control and leverage their capabilities. 

Single-cell level isotope probing is an orthogonal approach to molecular analysis. Unlike molecular techniques which capture potential activity, quantification of isotope uptake with secondary ion mass spectrometry (SIMS) directly reveals the record of metabolic activity by measuring isotope accumulation \cite{Pett-Ridge2012, mcclelland2020direct}. In addition, it offers exceptional sensitivity, detecting isotope ratio changes of less than a 1\% difference. While deployed in microbial ecology \cite{musat,gregorie,mcglynn} and in the analysis of nutrient shifts \cite{Kopf2015,Schreiber2016}, obtaining a granular view of metabolism at the single-cell level from isotope labeling data is challenging, since the measurement of a single isotope is an aggregate of metabolic fluxes, rather than a breakdown of specific pathways. 

How can single-cell isotope data be translated into an understanding of metabolic activity at the scale of individual pathways and fluxes? Traditional techniques for analyzing metabolic fluxes characterize cellular metabolic phenotypes in bulk by measuring fluxes through key pathways and combining mass spectrometry data with mathematical network modeling \cite{zamboni200913c,long2019high}. Here we integrate single-cell isotope data with statistical inference and metabolic modeling to infer granular details of metabolism at the single-cell level. Working with the cyanobacterial model organism {\em Synechocystis} sp. PCC6803, we find a population structure where cellular metabolic potential and growth rate is expanded by exudate cross-feeding. Employing single-cell level constraint-based metabolic flux analysis, we resolve individual call metabolism and its variability. 

\begin{figure*}
\begin{centering}
\includegraphics[width=1\textwidth]{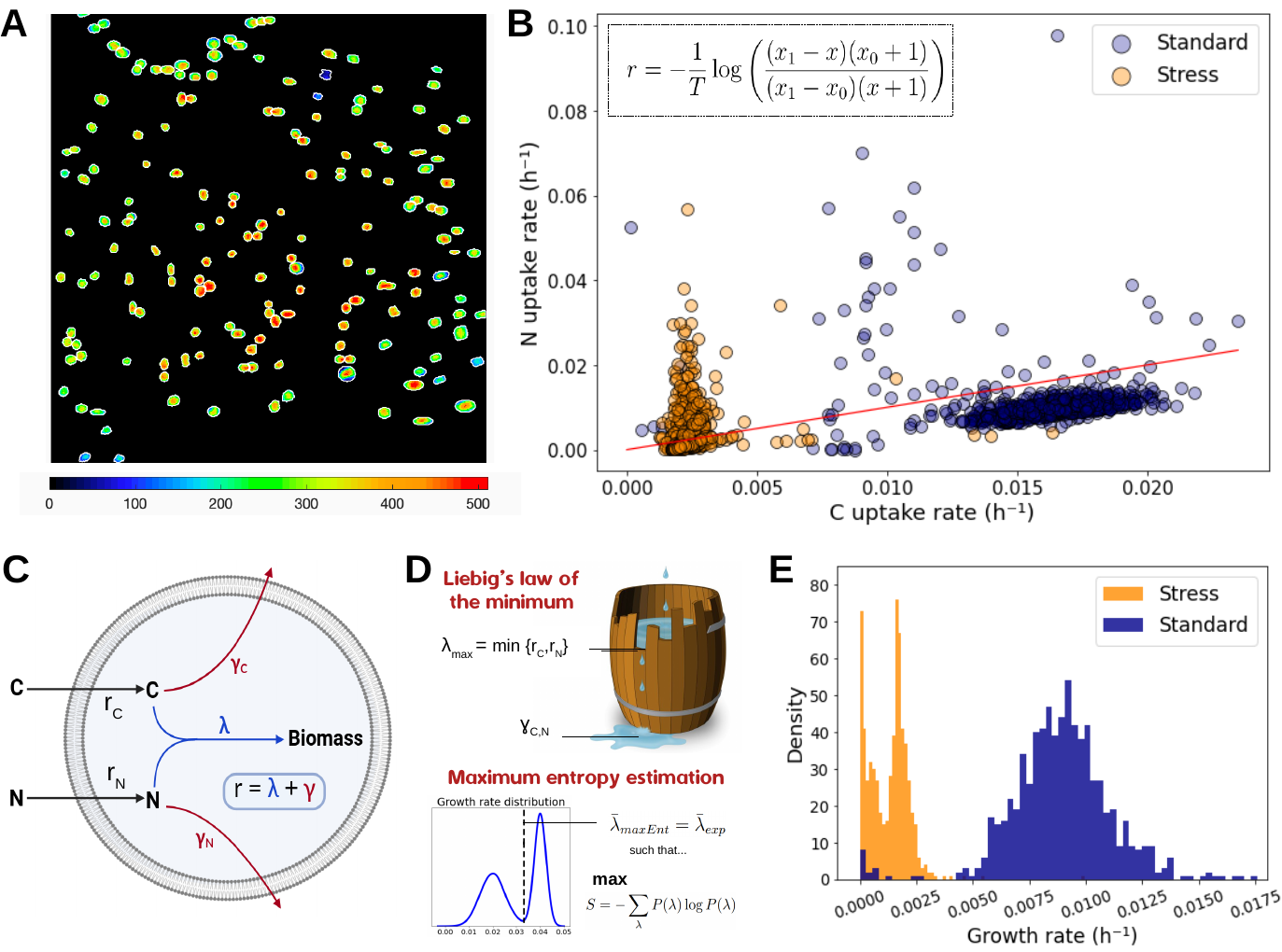}
\caption{\textbf{Using Isotope Labeling and SIMS to determine C and N uptake rates and growth rates.} 
\textbf{A.} SIMS image of cells deposited on an ITO glass coated slide displaying \textsuperscript{13}C ion counts for cells recovered at time T after isotopic labeling. Field of view: 100 x 100 µm; the color bar at the bottom represents ion counts. 
\textbf{B.} Individual cell biomass normalized C and N specific uptake rates under standard and stress conditions, calculated from isotopic ratios using the equation provided in the upper left corner which describes the exponential change in isotope ratios over time (SI, Sec. II). The red line ($y=x$) represents equal uptake rates in units of per hour.  \textbf{C.} Illustration of a simple mass-balance model for the uptake rates. The uptake rate $r$ of an element must equal the sum of the rate in which it is incorporated to the biomass (which is the growth rate $\lambda$) and the rate of turnover $\gamma$ due to an excess of atoms flowing out of the cell. \textbf{D.} Inferring the grow rate of individual cells with the maximum entropy approach. The probability distribution was calculated to maximize the entropy given two main constraints: experimental average growth rate of the bulk population and individual maximum growth rate derived from the Liebig's law of the minimum. \textbf{E.} Density plot showing the inferred distribution of the individual cell growth rates under standard and stressed conditions. } 
\label{fig1}
\end{centering}
\end{figure*}

\section*{Results and Discussion}


\subsection*{Metabolic-model-free estimation of cyanobacteria growth and turnover rates at the single-cell level}

We quantified carbon and nitrogen uptake at the single-cell level in the photoautotrophic cyanobacterium \textit{Synechocystis} sp. PCC6803, which had been maintained in the exponential phase through semi-continuous culture and spiked with \textsuperscript{13}C bicarbonate and \textsuperscript{15}N ammonium (Fig. \ref{fig1}A and SI Fig. S1) for 1/3 of a doubling time. This time frame covered both light and dark periods, meaning that the retrieved cells and metabolic states associated with them represent physiological processes integrated over incubation time. Our analyses cover several hundreds of cells in two growth conditions: one standard condition, and the other a nutrient stress condition with low oxygen (input gas changed from atmospheric to 80:20 N\textsubscript{2}:CO\textsubscript{2} 
, and sulfur (MgSO\textsubscript{4} reduced from 0.304 mM to 0.2 mM), and nitrogen (NH\textsubscript{4}Cl reduced from 17.65 mM to 0.1 mM). 

During balanced cell growth, the biomass composition remains constant during cell division. In this condition and in the absence of elemental turnover, elemental uptake rates are expected to be equal when normalized to biomass. These uptake rates would then fall along a line of slope 1 when multiple elements are plotted (Fig. \ref{fig1}B, red line). We observed that commensurate uptake of C and N for cells in our study population was rare: cells  predominantly fall above or below the line of equal uptake rate (Fig. \ref{fig1}B), suggesting that the SIMS derived uptake rate $r$ of each element has a biomass component (i.e., the growth rate $\lambda\geq 0$) and a turnover component corresponding to isotope enrichment decoupled from growth ($\gamma\geq 0$) (Fig. \ref{fig1}C), i.e. $r=\lambda + \gamma$. 

These element-dependent turnover rates generate uncertainty in the underlying single-cell phenotypes; even though we know the rate of change in isotope composition (uptake rate) from our experimental data, we cannot decompose them into their anabolic and turnover components. 
Most previous efforts devoted to calculate growth rates from SIMS isotopic labeling data have stopped at this point, referring for example to the rate as ``C-based specific division rate'' \cite{Berthelot2018}. Here we move further by making explicit the distinction between the uptake rate of an element and the actual growth rate of the cell.


Since the growth rate cannot exceed the minimum of either C and N elemental uptake rates 
we can use knowledge of the rates calculated from SIMS data to set the maximum growth rate of cell $i$ (SI, Sec. II) as
\begin{equation}
    \lambda_i^{\max}=\min\{r_{C,i},r_{N,i}\} ~~.
\end{equation}
This is inspired by Liebig's law of the minimum, but rather than focusing on exogenous nutrient levels, it emphasizes cellular acquisition which is quantitatively measured by the SIMS.

Since the growth rate of cell $i$ ($\lambda_i$) is now bound to lie between $0$ and $\lambda_i^{\max}$, we can quantify the likelihood of  $\lambda_i$ taking on a particular value using a probability distribution on the interval $[0,\lambda_i^{\max}]$. Without prior knowledge of the probability distribution, one should assume that all values are equally likely (a uniform distribution), meaning the expected growth rate of cell $i$, $\avg{\lambda_i}$, would be $\lambda_i^{\max}/2$. We can however remove some uncertainty by incorporating knowledge of the bulk growth rate from cell counting, $\lambda_{\mathrm{exp}}$, which provides an independent estimate of the population growth rate (SI, Sec. III). Specifically, we can impose that single-cell growth rates satisfy the condition
\begin{equation}
    \frac{1}{N}\sum_{i=1}^N\avg{\lambda_i}=\lambda_{\mathrm{exp}}~~,\label{2}
\end{equation}
which equates the population average as it is derived from individuals to the laboratory measured bulk growth rate. Note how in this setup individual cells are not independent, as the single cell growth rates are coupled to equal  the mean laboratory growth rate when averaged $\lambda_{\mathrm{exp}}$ in (\ref{2}). Statistical theory holds that the least biased distribution satisfying this requirement is the one that maximizes the Shannon entropy subject to a given mean \cite{Jaynes2003}. 
This results in the Boltzmann-Gibbs distribution, whereby the probability $P(\lambda_i)$ of observing a single-cell growth rate $\lambda_i$ is an exponential function of $\lambda_i$, (Fig. 1D)
\begin{equation}\label{hgjfdxghjyktj}
    P(\lambda_i)\sim \exp(\beta\lambda_i)~~~~~(0\leq \lambda_i\leq\lambda_i^{\max})~~.
\end{equation} 
Here $\beta$ is a parameter that ensures that individual cell growth rates will average to equal the population average in   (\ref{2}). $\beta$ can be found numerically by solving a non-linear equation that represents this matching condition and subsequently, the growth rate of each single-cell $\lambda_i$ can be analytically found to obtain the distribution of single-cell growth rates shown in Fig. \ref{fig1}E (see SI section IV for details on the estimation of $\beta$ and the growth rate distribution). 

In line with the heterogeneity observed in uptake rates (Fig. \ref{fig1}B), the inferred growth rates also exhibit substantial variability, with the highest growth rates occurring in the standard growth media (Fig. \ref{fig1}E and Fig. S2). The growth rate distribution in the standard media composition indicates that growth rates varies by ~20 fold in the population ranging from nearly 0 to 0.0175 h$^{-1}$. 
In contrast, under stressed conditions, the population appears to adopt a bimodal distribution, with distinct subpopulations exhibiting different growth rates. Specific to cyanobacteria, slow growth under nutrient stress agrees with previous reports of slower growth and chlorosis for nitrogen-starved cultures \cite{Richaud2001, Koch2019}; more broadly these results are also in line with an apparent shift to a bimodal growth rate distribution as doubling times increase, which was observed using single-cell isotopic measurements for a different group of microorganisms \cite{Kopf2015}. 





\subsection*{Maximum-entropy Estimation of Single-Cell Growth Rates are Consistent with Genome-Scale Metabolic Models}

\begin{figure*}
\begin{centering}
\includegraphics[width=1\textwidth]{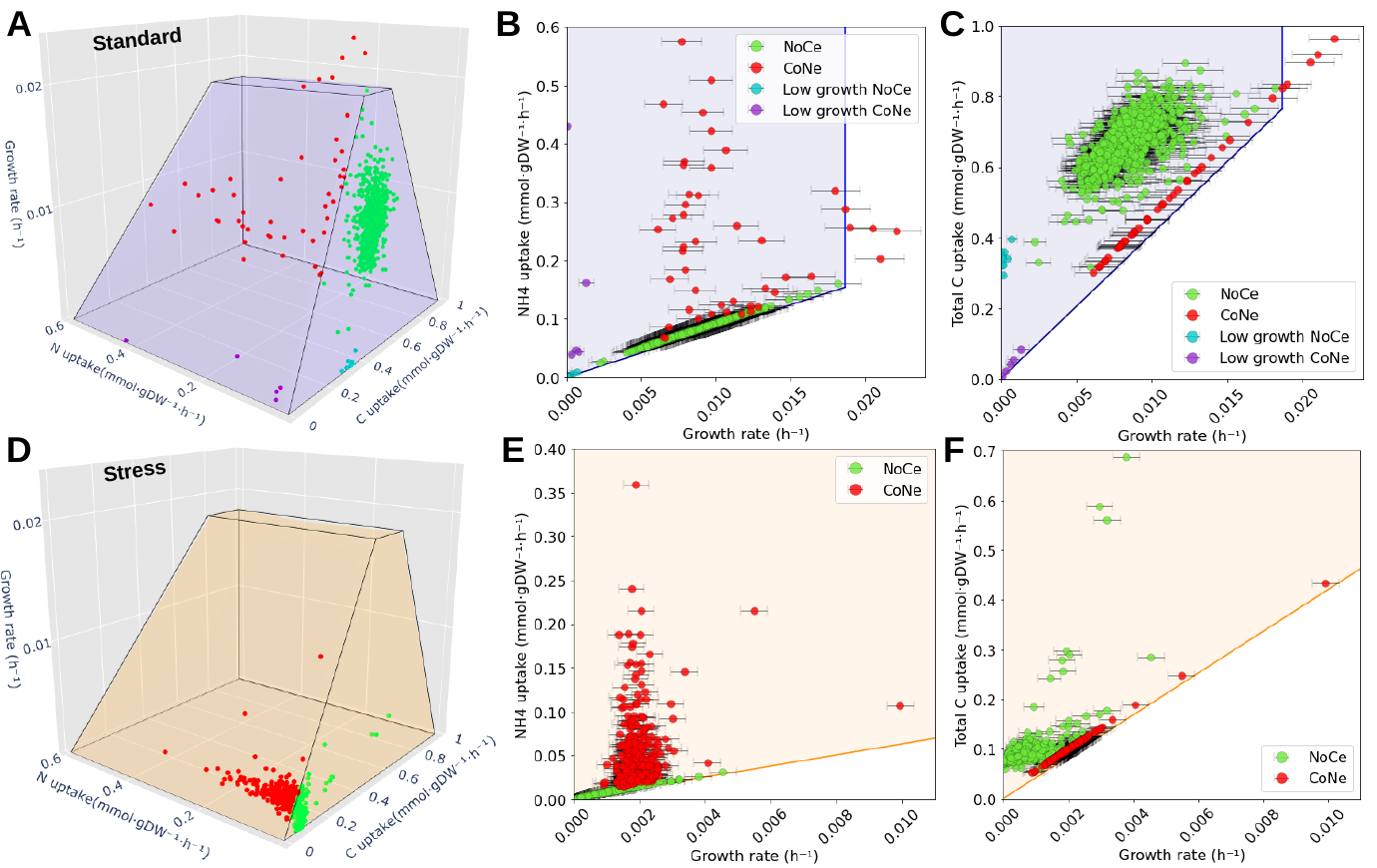}
\caption{\textbf{Inferred growth rates and uptake rates within metabolic boundaries. A.} 3D plot showing the gross uptake flux of C, N, and the growth rate distribution across the population under standard conditions. Each data point represents one cell, i.e. the individual growth rate inferred by the maximum entropy approach and N and C uptake rates derived from SIMS. The shaded area represents a low-dimensional projection of the feasible space of the high-dimensional genome scale metabolic model (see materials and methods and SI section V). Cells are colored based on clustering results. The cluster acronyms are as follows: CoNe — carbon optimal, nitrogen excess; and NoCe — nitrogen optimal, carbon excess. This clustering pattern is also observed with low-growth phenotype in standard conditions. \textbf{B and C.} 2D projections showing the relationship between growth rate and, respectively, nitrogen (B) and carbon (C) uptake fluxes. Shading and color coding are consistent with panel A. Note that the carbon (resp. nitrogen) axis represents the HCO$_3^-$ (NH$_4$) uptake flux within the model's feasible space, as HCO$_3^-$ (NH$_4$) is the sole carbon (nitrogen) source in the simulated medium.  \textbf{D, E and C.} Equivalent plots to A, B, and C, respectively, but showing data under stress conditions.} 
\label{fig2}
\end{centering}
\end{figure*}

Are the above inferred single-cell growth rates compatible with previous knowledge of this model organism?  Since {\em Synechocystis} sp. PCC6803 is a well studied organism with an available genome-scale metabolic model \cite{Nogales2012}, we compared the growth rates calculated from our single cell maximum entropy estimation with the growth rates permitted by this genome-scale metabolic model, treating growth in the presence of isotopes as an integrated period across light and dark periods \cite{sarkar2019diurnal}. The genome-scale model’s feasible space represents the single-cell flux configurations admissible in a given environment, 
which we calculated using Flux Balance Analysis (FBA, \cite{orth2010flux}). 

With the exception of 6 cells in the standard growth condition, the SIMS-derived uptake rates and inferred single-cell growth rates fit within a low-dimensional representation of the FBA model feasible space (Fig. \ref{fig2}; SI Sec. V for details). This demonstrates that the parameter estimation in the previous section -- which is metabolic model-independent and based only on the measurement of isotope uptake rates, statistical inference, and a constraint of Liebig's law -- yields physiologically meaningful results. Cells located along the boundaries of the FBA feasible space approximately maximize growth yields for carbon and nitrogen, suggesting the co-existence of states within the culture.

\subsection*{Cells Cluster into Metabolically Distinct Populations, Optimized for Either C or N Uptake}



We performed a clustering analysis to identify groups of metabolically similar cells in both the standard and stressed conditions. We used as input the information represented in the three axes of Fig. \ref{fig2}A and D, obtaining clusters of cells with similar metabolic phenotypes (see SI Sec. VI for details on the clustering protocol). In the standard condition, we identified four distinct clusters, categorized on two primary criteria: optimality for carbon or nitrogen utilization, and growth rate. This clustering was robust to a  1 percent random perturbation of single cell data, which is an error on the order of the SIMS measurement (SI Fig. 10).

The identified clusters are spread on axes of nitrogen uptake and growth rate in  Fig. \ref{fig2}B, with the shaded region representing FBA-defined maximum growth at the given nitrogen uptake rate. Cells in red are optimal for the carbon uptake and take up excess nitrogen (CoNe; Carbon optimal nitrogen excess), while cells in green are optimal for their uptake of nitrogen, yet take up excess carbon (NoCe; Nitrogen optimal, carbon excess). 
Although the CoNe and NoCe clusters are dominant in the population, these metabolic modes themselves appear stratified, with a subset of cells characterized by very low or negligible growth despite remaining metabolically active (low growth NoCe cells in purple and low growth CoNe cells in blue in the figure). 

In stressed conditions, the population shows slower growth, and the growth rate distribution is more compressed (see Fig. \ref{fig1}E and \ref{fig2}D-F). Differentiation is also compressed within this population, with 2 clusters identified instead of 4 in the standard condition. However, optimality of carbon and nitrogen utilization is observed in the same pattern as in the standard condition, with the CoNe subpopulation optimized for carbon and the NoCe cells optimized for nitrogen uptake. 




\begin{figure*}
\begin{centering}
\includegraphics[width=1.\textwidth]{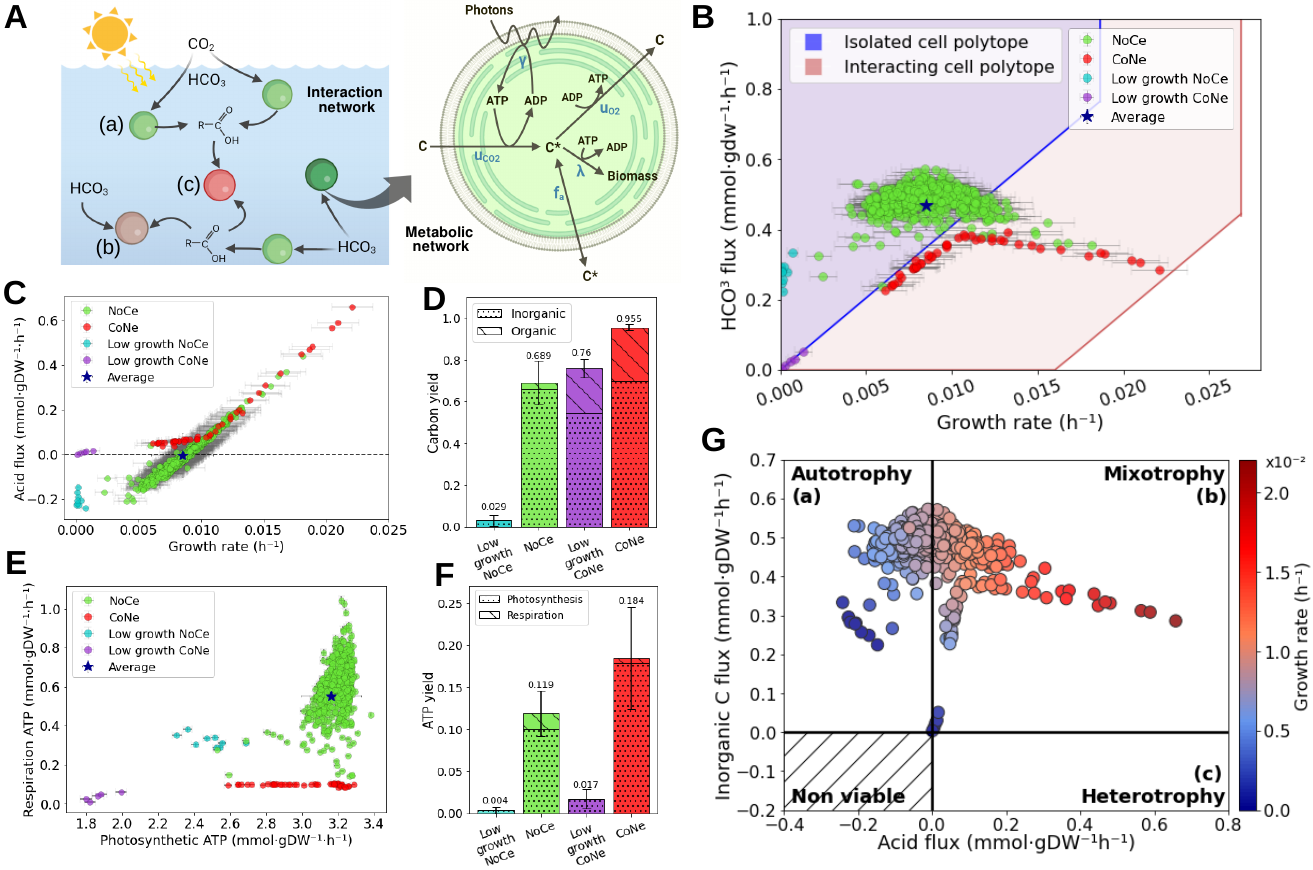}
\caption{ \textbf{Metabolic characterization of individual cells within an interacting population. A.} Illustration of a plausible scenario in which a population of metabolically diverse cells interacts through the exchange of fixed carbon, thereby forming an interaction network. (a) purely autotrophic cells that can secrete organic carbon, (b) mixotrophic cells that fix CO$_2$ while also consuming fixed carbon, and (c) purely heterotrophic cells that rely entirely on autotrophic excretion. Each cell within this network is represented by the simplified metabolic network on the right, consisting of only four independent fluxes, which ensures computational tractability. Positive fluxes represent intake, and negative fluxes excretion. \textbf{B.} Inorganic carbon uptake values obtained by sampling the solution space of an interacting cell for standard growth conditions. Shaded areas stand for the old and new polytopes, with and without interactions taken into account. Same color code as in Figure 2 is used for identifying the different clusters. \textbf{C.} Organic carbon flux against growth rate of individual cells within an interacting population. Positive and negative values indicate uptake and excretion respectively. \textbf{D.} Average carbon yield of each cluster and its proportion of inorganic and organic carbon contribution. \textbf{E.} Comparison of the flux through the two primary ATP synthesis modes—respiration and photosynthesis—at the single-cell level. \textbf{F.} Average ATP yield of each cluster and its proportion of photosynthesis and respiration contribution. \textbf{G.} Inorganic carbon uptake versus organic carbon flux for individual cells. The space is divided based on different plausible metabolic strategies, with cells represented as scattered points colored on a gradient reflecting their growth rates. 
}
\label{fig3}
\end{centering}
\end{figure*}

\subsection*{Constraining FBA with Maximum Entropy Modeling Allows Resolution of Individual Metabolic Fluxes and Predicts Organic Acid Secretion}
We next sought to clarify the metabolic differences between clusters observed in the standard growth condition. Since each cell is associated with knowledge of growth rate and nitrogen and inorganic carbon uptake rates, we can obtain a quantitative picture of the metabolic fluxes through key pathways at single-cell resolution by further constraining the genome-scale metabolic model with these values. The feasible space of the model is defined by the conditions
\begin{gather}
\mathcal{S}\cdot \bf{f}_i  =0~~, \\
\bf{f}_i \in [f^{m},f^M ]~~, \\
\bf{u}_i \in [u_{i,exp}-\delta_{u_{i,exp}},u_{i,exp}+\delta_{u_{i,exp}}]~~.
\end{gather}
Equation 4 represents the steady-state condition, where ${\cal S}$ is the stoichiometric matrix and $\bf{f}$ denotes the vector of reaction fluxes. Each flux is bounded between a specified minimum ($\bf{f}^m$) and maximum ($\bf{f}^M$) value. Uptake fluxes ($\bf{u}_i$) that have been experimentally measured at the single-cell level are fixed to their corresponding mean values, with associated standard deviations used to account for measurement uncertainty (the same applies to the growth rate). To inject empirical information, we assigned equal probability to all flux configurations that align with the observed single-cell growth and uptake rates. 
Feasible single-cell flux vectors thus span a high-dimensional space with flat boundaries (i.e. a polytope, see Materials and Methods).   
To efficiently sample this space, we resorted to Monte Carlo methods optimized
to derive distributions of individual metabolic fluxes for cells across the population (see \cite{narayanankutty2025metabolic} and SI Sec. IX).

Using this approach, we obtained single-cell estimates for all 865 reaction fluxes represented in the genome-scale metabolic model. Notably, 78\% of the network was well defined and resolved with a relative error (standard deviation divided by the mean) below 30\%. Despite a substantial portion of the network being inactive, around 40\% of the network was non-zero flux and well-determined. These reactions were mainly associated with metabolic functions such as carbon fixation, maintenance and aminoacid metabolism (see SI Fig. S11). 

NoCe cells that maximize growth yield on ammonium but excessively fix carbon are predicted to excrete organic acids (SI Figure S4). Although organic acid exudation is a known energy regulation mechanism \cite{Cano2018}, predicted secretion fluxes could result in unreasonably low pH levels in the medium (SI Sec. VII and Fig. S5). This acidification motivates a hypothesis where a fraction of the population could take up excreted acids, alleviating acidification of the medium and contribute to biomass production.

\subsection*{Allowing Cell Interaction Expands the Metabolic Solution Space to Accommodate All Cells}

It has previously been reported that \textit{Synechocystis} sp. PCC6803 uses organic compounds, including acids, for mixotrophic growth \cite{haavisto2024high}. 
Moreover, it was recently reported that cross-feeding between \textit{Prochlorococcus} and \textit{ Synechococcus} may take place via carbon nanotubes \cite{Angulo-Cnovas2024}. We thus generalized our model to allow for the intercellular exchange of organic acids to test the idea that reciprocal use patterns of carbon and nitrogen reflect a division of metabolic labor and can expand the metabolic capabilities of the population. 

Considering carbon exudation and uptake in a population, we can envision a population consisting of (a) purely autotrophic cells that can secrete organic carbon, (b) mixotrophic cells that fix CO$_2$ while also consuming fixed carbon, and (c) purely heterotrophic cells that rely entirely on autotrophic excretion (Fig. \ref{fig3}A). 
Could a nitrogenous compound also be exchanged? Such exchanges have been observed in cyanobacteria \cite{braakman2025global} and might be considered here based on the observation of the nitrogen-excessive CoNe cluster. However, since nitrogen exchange typically occurs via organic molecules, this hypothesis is inconsistent with the observed carbon uptake in this cluster, which is near the minimum required to support growth. In other words, these cells do not acquire enough carbon to allow for its excretion alongside a nitrogenous compound. It is possible that an inorganic nitrogen compound could be exchanged, but since we do not have experimental evidence for this we do not consider the possibility further here. 

Capturing exchanges between multiple cells within the population in the genome-scale model requires the simultaneous computation of hundreds of fluxes, which is computationally infeasible. Therefore, in order to reduce the search space, we preliminarily attempted to derive additional constraints by studying a highly coarse-grained version of the genome-scale model. In particular, we developed a simplified model that describes four independent fluxes per cell: carbon uptake and excretion (partitioned into organic and inorganic forms), biomass synthesis, and energy generation via photosynthesis and respiration (Fig. \ref{fig3}A and SI Sec. VIII). This framework captures the carbon and energy balance of a typical cyanobacterium.
We added to this simple single-cell constraint-based model intercellular diffusion constraints that enable interactions through exchange along the lines of \cite{narayanankutty2025metabolic} (see SI Sec. VIII, equations (49), that couple the uptakes of single cells).
We then studied the feasible space of a population of such cells, each represented by one such reduced metabolic model randomly placed in a three-dimensional space matching the experimental cell density (SI Sec. IX). Unlike the previous section's approach, which computed each cell metabolism independently and excluded interactions between cells, this model enables the simultaneous calculation of these four single-cell fluxes across the population in a computationally efficient manner, allowing a derivation of carbon exchange through an interaction network (SI Fig. S7).

The first key result is that the permission of interactions between cells expands the metabolic solution space, enabling faster growth rates and accommodating all measured cells in the polytope volume (see Fig. \ref{fig3}B, pink-shaded area vs. purple-shaded area). Compared to initial model inference results where interactions between cells were not permitted (Fig. \ref{fig2}C), cells in the model allowing interactions generally take up less inorganic carbon: acids secreted by a subset of the population are taken up by a different cluster of the population.
The population also includes cells with no acid flux that fix carbon with complete uptake efficiency i.e. no exudation (Fig. \ref{fig3}C). Interestingly, the average cell displays no acid flux and has complete carbon uptake efficiency, indicating minimal or negligible acid accumulation in the growth medium—consistent, with previous findings under these experimental conditions and an estimation of pH during the growth  (SI Fig. S5).

\subsection*{Cells are Realized Only as Autotrophs and Mixotrophs; All Cells Use Light as Energy Source}

Cells that achieve higher growth rates than their purely autotrophic counterparts adopt a mixotrophic lifestyle, with growth rates increasing in proportion to acid uptake (Fig. \ref{fig3}C).  Looking at Fig. \ref{fig3}B and C together shows that beyond a growth rate  $\sim\,$0.01 h$^{-1}$, inorganic carbon fixation plateaus, with further growth being sustained by the uptake of additional carbon in organic form (Fig. \ref{fig3}C).
Furthermore, the fastest growing cells derive more carbon from organic acids than from CO$_2$, while the slowest growing metabolically active cells (excluding the slowly growing cells of the purple cluster) excrete acids, in line with a previous observation of a higher release of organic carbon at lower growth rates \cite{Dubinsky2001}. Within the population, there appear to be no purely heterotrophic cells (Fig. \ref{fig3}G), although such a phenotype (cell of type (c) in Fig 3A) could be in principle viable.

Insights into the energetics of individual cells within the interacting population can also be derived. ATP in these cells is primarily generated through two mechanisms: phototrophically, using light, or chemotrophically, via the oxidation of organic compounds. Fig. \ref{fig3}E illustrates how clusters segregate based on these two ATP synthesis strategies. A key distinction between CoNe and the NoCe cells lies in their energy source: while NoCe cells appear to have lower carbon and ATP yields, CoNe fully benefit from organic carbon - since they still derive energy from light, they can allocate all available organic carbon towards biomass synthesis without incurring losses through respiratory ATP production or using energy to pump and fix higher amounts of inorganic carbon. 

Notably, all cells utilize light as an energy source, while respiration serves as an additional pathway predominantly in populations exhibiting overflow metabolism (NoCe clusters). This indicates that a portion of the organic carbon is not secreted but instead respired, releasing CO$_2$ back into the environment. 
Cells which overfix carbon and excrete it also derive a greater fraction of their ATP through this respiration in turn performing a futile cycle, perhaps as an anticipatory strategy to the day-night switch \cite{newsholme1984substrate}.

\begin{figure*}
\begin{centering} \includegraphics[width=1\textwidth]{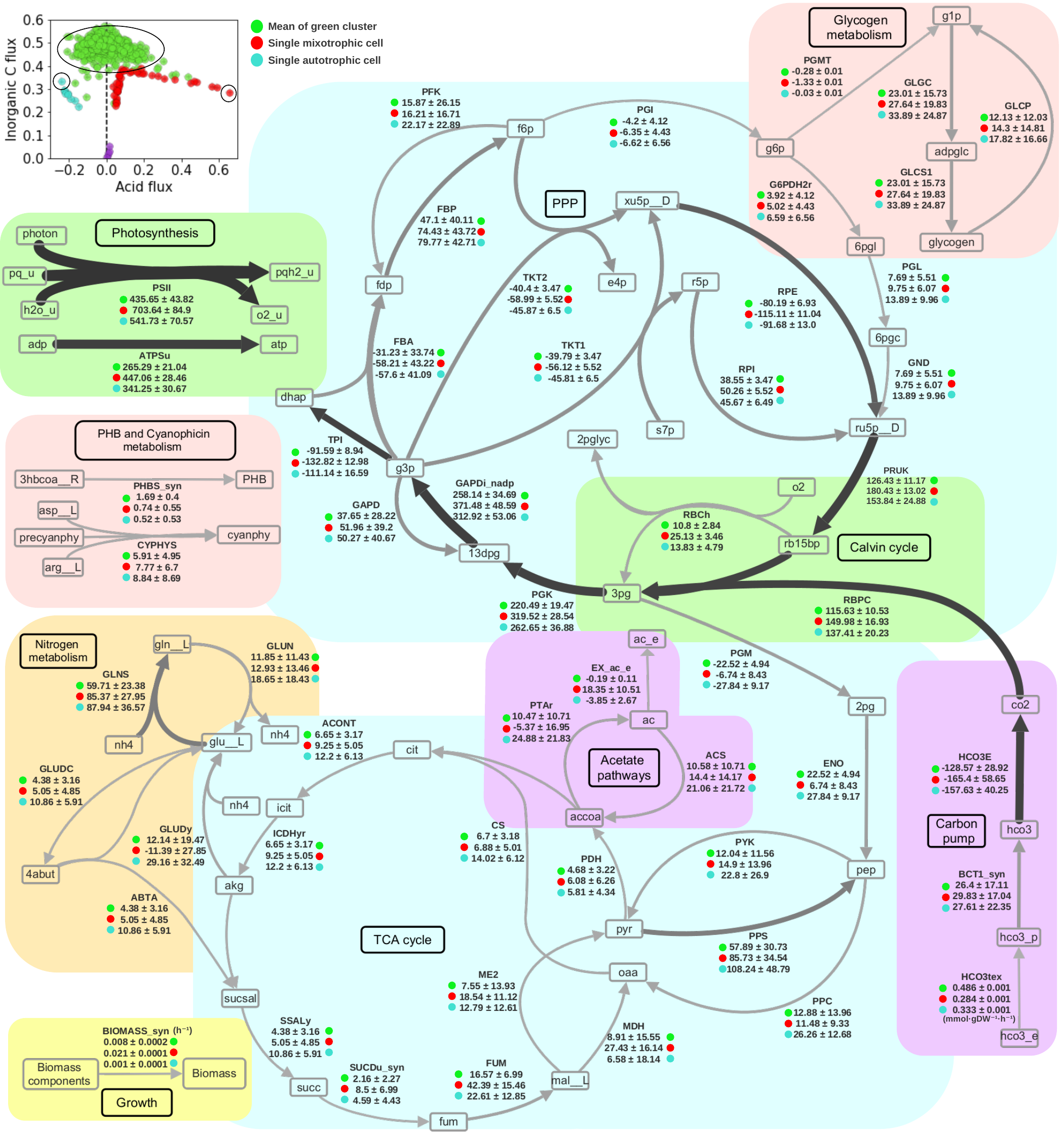}
\caption{ \textbf{Single cell metabolic flux analysis.} With four fluxes available per cell (N and C uptake rates from SIMS; inferred growth rate and acid exchange), metabolic fluxes can be estimated. Three values are shown for each flux in the metabolic map and correspond to cellular states shown in the inset at the top left: (1) the average and standard deviation of the flux relative to HCO\textsubscript{3}\textsuperscript{-} intake across NoCe cells, (2) a single CoNe cell with the highest acid influx (far right and red in the inset), and (3) a single NoCe low-growth cell with the highest acid influx (far left, blue in the inset). Arrow thickness is proportional to the relative flux intensity of the green cluster. All values are dimensionless, except for the HCO\textsubscript{3}\textsuperscript{-} uptake flux and the growth rate, which are reported in mmol·gDW\textsuperscript{-1}·h\textsuperscript{-1} and h\textsuperscript{-1}, respectively. Acronyms for compounds and reactions are taken from the genome-scale metabolic model \cite{Nogales2012}.
}
\label{fig4}
\end{centering}
\end{figure*}

\begin{figure*}
\begin{centering}
\includegraphics[width=1.\textwidth]{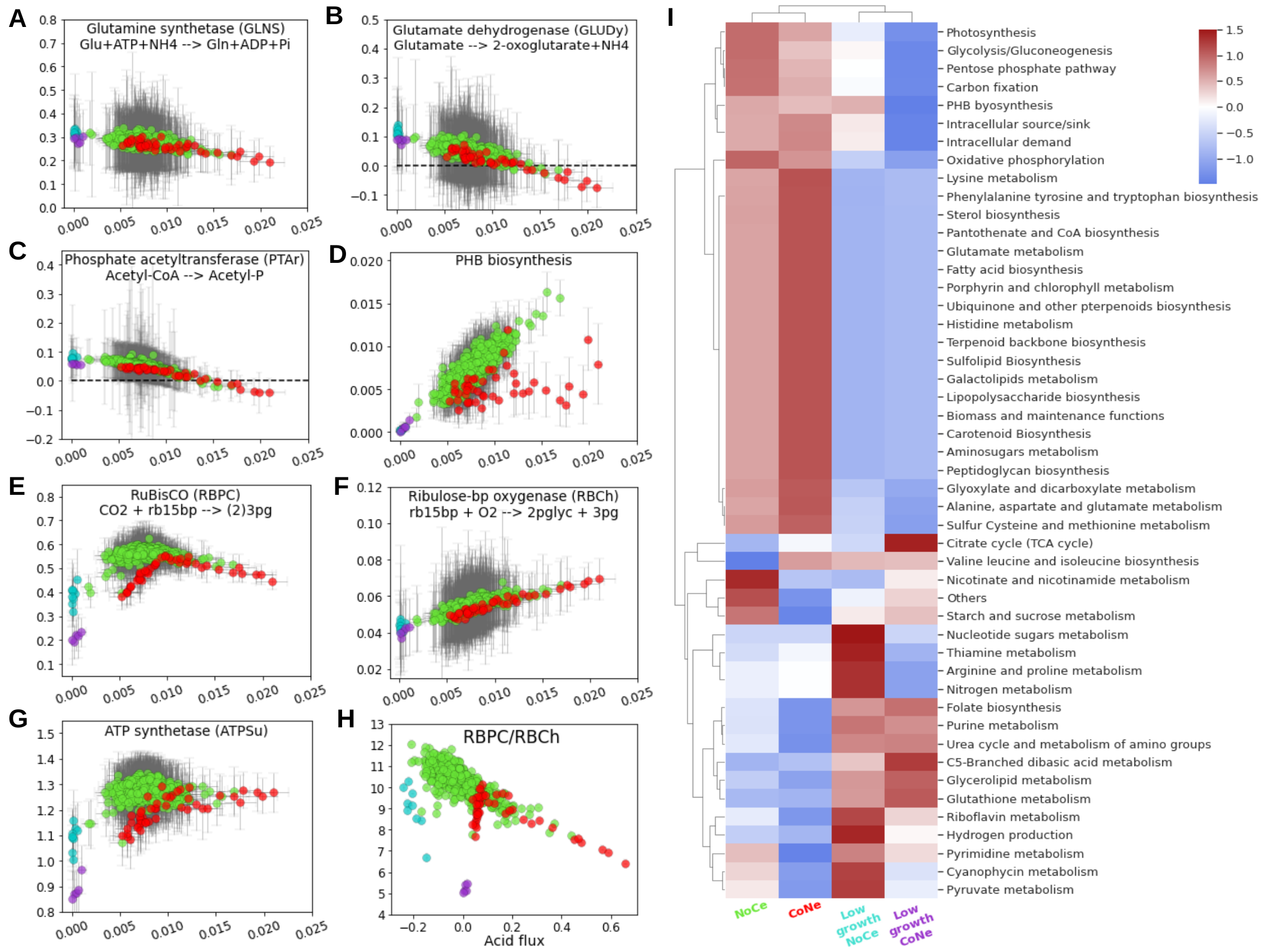}
\caption{ \textbf{Metabolic heterogeneity among single cells and clusters in the interacting population.} \textbf{A-G. } Scatterplots of single-cell  metabolic fluxes (mmol·gDW$^{-1}$·h$^{-1}$) on the y-axes vs. growth rates (h$^{-1}$) on the x-axes. \textbf{H. } Scatterplot of Rubisco flux of the carboxylase reaction/oxygenase reaction (Rubisco specificity) vs. the total acid flux. \textbf{I. } Clustermap showing the results of a metabolic pathway enrichment for the different clusters. The activity of each pathway was calculated as the sum of the absolute flux through every reaction associated to it (see Methods section). Values are normalized by Z-score in the rows. 
}
\label{fig5}
\end{centering}
\end{figure*}

\subsection*{Single-cell metabolic flux analysis}

To further characterize the internal metabolism of individual cells within the interacting population, we repeated the metabolic flux analysis at the single-cell level, this time incorporating the estimated fluxes of photon uptake, respiration, and the partitioning of carbon into inorganic and organic forms from the simplified model described above. After constraining the corresponding reactions in the model, which now include organic acid uptake and excretion, we resampled the entire metabolic network, obtaining flux estimations for every internal reaction. 

Our flux analysis reveals significant variability across the identified clusters in several intracellular and exchange reactions that were not directly targeted with isotope labels. Fig. \ref{fig4} presents a map of the central metabolism of cyanobacteria that highlights the main pathways, including the reductive pentose phosphate pathway (Calvin-Benson cycle), glycolysis, the pentose phosphate pathway, and the TCA (Krebs) cycle, along with the inferred reaction fluxes which show the diversity of metabolism observed in the population. This representation highlights global variability in metabolic fluxes which we infer at the single cell level.


While Figure 4 provides a graphical overview of metabolism with representative fluxes for cell-members, flux variability can also be plotted for every cell in the population Fig. \ref{fig5} A-G. 

The NoCe subpopulation optimizing nitrogen yield is the numerically the most prevalent in the population. In \textit{E. coli}, nitrogen assimilation via ammonia occurs through the GLNS and GLUDy pathways, regulated by the two-component system NtrBC \cite{Doranga2024}. In contrast, \textit{Synechocystis} relies solely on the more energy-intensive GLNS containing GS-GOGAT pathway, which has inducible isoforms \cite{Inabe2021}. The limited ammonia uptake in these cells aligns with their inability to utilize the alternative amination of alpha-ketoglutarate. We speculate that differential expression of these isoforms could explain the nitrogen uptake differences observed among the clusters. This nitrogen-optimized subpopulation appears to be the primary source of organic acid spillover, which might result from the use of CO2 as an electron acceptor to regenerate reduced cofactors, akin to overflow metabolism in \textit{E. coli} \cite{Vemuri2006} and observations of another photosynthetic bacterium \cite{McKinlay2010}. 


Glutamine synthetase (GLNS) has previously been shown to have a 2-4 fold higher expression (glnA) in nitrogen starved cells \cite{Reyes1994}, which is consistent with the NoCe low growth cluster showing the highest relative glutamine production, almost twice that of the CoNe and NoCe clusters (\ref{fig5}A). Similarly, glutamate dehydrogenase (GLUDY) activity is the highest in the slowly growing blue cluster (Fig. \ref{fig4} and \ref{fig5} B): it is known to have high NADP dependence and being important to sustain growth in non-exponential phases \cite{Chavez1999}.

The second-largest cluster, CoNE is optimized for carbon yield and provides insight into this regulation. Specifically, in the case of acetate, we observed up-regulation of the acetyl-CoA synthetase (ACS) pathway and a reversal of the ACK-PTAr pathway, in contrast to its direction in the nitrogen-optimized cluster (acetate pathways module of Fig. \ref{fig4} and Fig. \ref{fig5}C). Acetate is secreted in the NoCe clusters, which is in direct relation to Phosphotransacetylase (PTAr) activity \cite{Roussou2025} (Fig. \ref{fig5}C). Other organic acids (pyruvate, fumarate, succinate, citrate, alpha-ketoglutarate, and malate) show similar release patterns SI Fig. S8). This suggests a regulatory mechanism similar to that found in \textit{E. coli}, where recent evidence indicates the coexistence of acetate-producing and acetate-consuming phenotypes \cite{enjalbert2017acetate, millard2021control}. However, unlike in \textit{E. coli}, where acetate-feeding cells grow more slowly, here the cells utilizing fixed organic acids are actually faster-growing mixotrophs, achieving growth rates paradoxically higher than those predicted by flux balance analysis.

The synthesis of Polyhydroxybutyrate (PHB), supposedly a carbon and energy storage metabolite, is linearly correlated with growth rate in the NoCe cluster, and shows high heterogeneity within the CoNe cluster (Fig. \ref{fig5}D), which is in agreement with previous findings on PHB phenotypic heterogeneity where varying PHB content was found in single cells \cite{Koch2019, Koch2020, Koch2020b}. 

The Rubisco fluxes are well-constrained, with an average single-cell relative error of roughly $10\%$. Rubisco activity is related to the population clusters and we can compare the fluxes of RuBisCO carboxylase reaction (RBPC) (Fig. \ref{fig5}E) to the oxygenase reaction (RBCh) (Fig. \ref{fig5}D) and consider the flux ratio (Fig. \ref{fig5}H). Since the carboxylase-to-oxygenase activity ratio depends on substrate availability (CO$_2$ vs. O$_2$)\cite{Flamholz2019}, it reflects the local cellular microenvironment and carboxysome state, which can now be resolved at the single-cell level. In agreement with previous findings on the impact of pH on Rubisco efficiency regulation \cite{mangan2016ph}, we found a strikingly simple linear relation between this ratio and the total acid uptake. 

The NoCe cluster (green) carries the highest Rubisco specificity of carboxylase to oxygenase reaction at intermediate growth rates ((Fig. \ref{fig5}H) and SI Fig. S9, panel B); the low growth NoCe cluster in blue has intermediate Rubisco flux ratios; and the low growth CoNe cluster in purple has low Rubisco flux ratios. Meanwhile, the CoNe cluster exhibits a piecewise linear trend: Rubisco flux ratio increases with growth rate and acid acid flux up to a threshold approximately coinciding with the average population growth rate, after which it declines. Interestingly, Rubisco’s oxygenase activity (RBCh) follows a different trend (Fig. \ref{fig5}F), increasing monotonically with growth rate, independent of cluster identity. 

Similarly, enolase activity increases with growth rate in the CoNe cluster roughly until the average population growth rate, after which it decreases (see SI Fig S9, panel A). Enolase catalyzes the conversion of 2-PG to PEP in glycolysis, and regulatory changes happen in response to changes in carbon partitioning inside the cell either into isoprenoids or glycerolipids \cite{knowles2003genome,Polle2014}. Enolase activity is highest in the NoCe cluster and decreases at higher growth rates.
Cells in this cluster are growing at the edge of feasible nitrogen incorporation space - and previous work has shown that glycerolipid content decreases in nitrogen starved cells and the increases when it becomes available again \cite{Kobayashi2020}.

We are also in a position to refine previous observations of mixed growth modes in the culture. Specifically, mixotrophic cells in the CoNe cluster exhibit photoheterotrophic growth, where energy is fully derived from photosynthesis, while assimilated organic carbon is directed entirely toward biomass synthesis. This is consistent with the higher Rubisco fluxes reported in earlier studies \cite{McKinlay2010}.

Finally, ATP generation through ATP synthetase (ATPSu) shows little variation with increasing growth rates in the NoCe and CoNe clusters, and greater variability with slow growth in the purple and blue clusters (Fig. \ref{fig5}G). Previous studies have identified the higher use of NADPH and ATP under autotrophic and mixotrophic conditions and a higher ATP/NADPH ratio in mixotrophic growth compared to autotrophic growth \cite{Kugler2023}.


In the metabolic model, each reaction is associated with a specific subsystem—representing a metabolic pathway or function—allowing us to perform a pathway enrichment analysis (SI Sec. X). The results, visualized as a clustermap in Fig. \ref{fig5}I, highlight the differential activity of metabolic pathways across the four clusters. 

As expected, processes such as photosynthesis and carbon fixation are most active in the NoCe cluster which over-fixes CO$_2$. Additionally, anabolic biosynthetic functions are more pronounced in the two largest clusters, with the CoNe cluster exhibiting the highest activity, consistent with its elevated biomass synthesis rates. 
In contrast, the low growth NoCe and CoNe clusters display low rates of photosynthesis, carbon fixation, and many anabolic functions. However, some biosynthetic pathways—particularly those related to amino acid and nucleotide metabolism—show increased fluxes in these clusters, suggesting that although these cells are not efficiently growing, they remain metabolically active. 
Notably, the NoCe clusters exhibit the highest activity in the biosynthesis of cyanophycin, a nitrogen storage compound, perhaps reflecting life on the "nitrogen edge" of growth.

\section*{Conclusions}

Single-cell studies are reshaping our understanding of microbial populations, revealing that genetically identical cells can exhibit distinct phenotypes. Intracellular molecular processes are inherently stochastic, leading to phenotypic heterogeneity even in fundamental metabolic traits \cite{kiviet2014stochasticity}. This variability may serve as a bet-hedging strategy, improving population survival in fluctuating environments \cite{Ackermann2013,Ackermann2015}. Metabolic differentiation seems to be common, as seen in division of labor within \textit{Anabaena} cysts or sporulation in \textit{Caulobacter} \cite{popa2007carbon, poindexter1964biological}, and also in single-cell transcriptomics studies which document variability in gene expression \cite{taniguchi2010quantifying, Dar2021, rosenthal2018}. A comprehensive understanding of metabolic phenotypes is essential for advancing microbial ecology and optimizing biotechnological applications. However, previous studies have been constrained by their focus on specific compounds or genes, limiting a broader perspective on metabolic variability. 



Genome-scale metabolic models account for the metabolic capabilities of an isolated cell in a particular environment dictated by the constraints on exchange reactions. At the population level, this translates to assuming a homogeneous population. In reality, however, cells exist within populations where interactions can occur among individuals (from one or multiple species), thus modifying the environment and growth constraints and making a single-cell approach necessary. 

Our findings suggest that intercellular metabolite exchange plays a crucial role in shaping population-level metabolic dynamics. We find that interactions can expand metabolic capabilities of the organism studied, effectively increasing the feasible solution space of theoretical models. The presence of carbon-excreting cells and mixotrophic individuals highlights a more complex metabolic landscape than previously assumed. By incorporating acid exchange into our metabolic model, excessive acid secretion was reconciled and high growth rates accommodated. These interactions point to a dynamic community where autotrophic and mixotrophic growth coexist, potentially mimicking natural microbial ecosystems where nutrient sharing fosters survival under fluctuating conditions.

Our results are in line with empirical evidence about metabolic coupling in cyanobacterial communities in the natural environment, where as primary producers, they play an important role in providing organics to heterotrophic communities \cite{bateson1988photoexcretion, stuart2016cyanobacterial, Eigemann2022}. It is also known that some primary producers such as Prochlorococcus rely on organic carbon to a greater extent than carbon fixation \cite{Wu2022}. At the same time, inter-species interactions play a crucial role in diversity \cite{Kost2023} and possibly chemotaxis \cite{Paerl1985} at the ecosystem level. 

Our single-cell metabolic flux analysis provides insights into key enzymatic pathways, such as Rubisco activity and nitrogen assimilation. Our results indicate that distinct clusters exhibit different metabolic strategies: the energetically costly nitrogen assimilation pathway employed by \textit{Synechocystis} may be responsible for the observation that most cells optimize nitrogen yield (NoCe), while others prioritize carbon turnover (CoNe). Enzymes within the cell do not operate in isolation, and the observed correlation between acid uptake and Rubisco efficiency suggests our flux estimates relate to the intracellular state and a regulatory mechanism influenced by  pH, supporting previous findings on the impact of metabolic byproducts on photosynthetic efficiency \cite{mangan2016ph}. 

One limitation of this study consists of the uncertainty in flux estimation for many pathways, including the rate of inter-cellular metabolic exchanges. To address this, the approach of integrating single-cell measurements with other bulk data using the maximum-entropy principle could be expanded beyond growth rate estimation (see also \cite{de2016growth,de2018statistical,fernandez2019maximum,tourigny2020dynamic,rivas2020metabolic,preciat2021mechanistic}). For example, incorporating bulk mass spectrometry and exudation rate measurements — such as those obtained through liquid chromatography — could provide additional constraints on flux averages, similar to traditional metabolic flux analysis. Combining these bulk measurements with single-cell data would enhance the resolution of metabolic exchanges and reaction fluxes, offering a more detailed understanding of cellular metabolism at single cell resolution.

Our results challenge the mechanistic view —often adopted in computational studies such as Flux Balance Analysis — that bacterial metabolic fluxes are precisely tuned to optimize an objective function, such as the growth rate/yield  \cite{bruggeman2020searching}. Instead, we reveal a more complex yet coherent picture, where optimality principles still play a role within the framework of true cellular ecology. Based on these findings, we can state that metabolic interactions within populations, rather than single-cell optimization, may shape overall carbon fluxes instead. Perhaps surprisingly, when such interactions are accounted for, empirical inter-cellular variability that could have been generically ascribed to `noise' or `randomness' (e.g., in gene expression or metabolism) takes on a coherent, structured organization, suggesting more intricate metabolic strategies at play. 

Cyanobacteria have been vital in global biogeochemical cycles for billions of years \cite{davin2025geological}, thriving in marine, freshwater, and terrestrial environments \cite{Whitton1992}. Their abundance and persistence make them an important target for further studies of their metabolic heterogeneity, especially in fluctuating and nutrient limited environments. The single-cell variability in metabolic fluxes we predict suggests an astonishing latent capacity for CO$_2$ fixation, where even a twofold variation in this process could impact atmospheric CO$_2$ levels on a scale comparable to anthropogenic emissions \cite{sarmiento1984new}. 




\section*{Materials and Methods}

\textit{For a graphical summary of the workflow see SI Fig. S12}
\subsection*{Cultivation and growth monitoring}

The freshwater cyanobacterium \textit{Synechocystis} sp. PCC6803 was cultivated in two growth conditions. 
In the first condition, standard freshwater BG11 medium \cite{Stanier1971} substituting NaNO$_3$ with NH$_4$Cl was used to grow the cyanobacteria with a modern atmosphere headspace. 
In the second condition the medium was modified to mimic nutrient limiting conditions (after \cite{Anbar2002, Farquhar2010, Saito2003}) and the culture was grown in a gas phase of N$_2$ (80\%), CO$_2$ (19.1\%) and O$_2$ (0.986\%). Here MgSO$_4$ concentrations were lowered from 0.304 mM to 0.200 mM and NH$_4$Cl was limited from 17.65 mM to 0.1 mM.
Cultures were incubated at 21° C with a 12:12 h light:dark cycle with a light intensity of \SI{50}{\micro\einstein} during the illumination phase. 
11.3 ml of 70 ml culture in standard and 2.62 ml of 70 ml culture in limited growth conditions were replaced with fresh medium every 24 h to maintain the cultures in semi continuous states in the exponential phases of their growth cycles.

\subsection*{Stable Isotope Labeling}

To quantify the substrate uptake in the different conditions, stable isotope labeling experiments were conducted in biological duplicates. 
\textsuperscript{13}C bicarbonate (NaH\textsuperscript{13}CO$_3$) and \textsuperscript{15}N ammonium (\textsuperscript{15}NH$_4$Cl) were used for assimilation analyses. 
\textsuperscript{13}C and \textsuperscript{15}N concentrations were chosen to be ~10 x higher than their natural abundances and were finalized at 5\% for \textsuperscript{15}N and 10\% for \textsuperscript{13}C. 
10 ml of cells were taken from the original cultures 1 h after the start of the 12 h light cycle, centrifuged and resuspended in duplicate in 30 ml airtight glass vials with 10 ml isotopically enriched media. 
The cultures were then grown and samples were taken after 21 h in the standard condition and 90 h in the stress condition, corresponding to approximately one third of a growth cycle in both conditions. 
0.5 ml of liquid culture were sampled from each vial. 
Due to the high cell density, 1 ml of a 1:5 dilution for the standard and 1 ml of a 1:2 dilution for the stress condition were filtered onto a \SI{0.2}{\micro\meter}  PTFE filter, which was washed twice with 0.2 ml miliQ water. 
The filters were then transferred with the cell-containing side in contact with indium tin oxide coated microscope slides and left to dry overnight at room temperature.
Afterwards the filters were removed from the slides and the slides were stored airtight until SIMS analyses.


\subsection*{Secondary ion mass spectrometry}

The instrument used for this analysis is a Cameca IMS 7f-GEO (Cameca, France) located at Washington University St Louis, MO, USA. 

Measurements were carried out in \SI{100}{\micro\meter} x \SI{100}{\micro\meter} areas and for each condition and biological replicate enough areas were analyzed to yield several hundred cells. 
For each measurement, 100 planes were analyzed with a primary ion beam current of ~20pA. 

The standard error for \textsuperscript{12}C/\textsuperscript{13}C and \textsuperscript{14}N/\textsuperscript{15}N was 1 – 2\%. 
SIMS data was processed in WinImage: planes were accumulated, cells manually identified as regions of interest and isotope ratios calculated within ROIs for the accumulated masses. 
To obtain the \% deviation of the corrected measured isotopic ratio from its natural abundance, archival natural abundances were used as the reference value for C and N.

\subsection*{Constraint based metabolic network modeling}

The exhaustively curated genome-scale metabolic model of \textit{Synechocytis} sp. PCC6803 (iJN678) published by \cite{Nogales2012} served as the basis for the metabolic modeling. The network comprises 795 metabolites and 865 reactions, divided in 55 metabolic pathways across 4 compartments (cytosol, periplasm, thylakoid and extracellular space). Boundary fluxes were fixed to simulate experimental conditions as described in SI Sec. VI. Flux balance analysis (FBA) \cite{Orth2010} was applied to calculate the maximum growth rate using the COBRA toolbox \cite{Ebrahim2013}. In both standard and stressed conditions $\lambda_{max}\sim 0.018$ h$^{-1}$, indicating that the limiting factor is light. To calculate the bounds defining the feasible space, the growth rate was fixed at different values iterating between 0 and the maximum growth rate, while the objective function was set to be either the carbon or the nitrogen uptake. At each value of the growth rate, FBA was employed to calculate the maximum and minimum values for the uptakes. In order to compute internal fluxes of metabolism, a single-cell metabolic flux analysis was performed. The sampling algorithm implemented in COBRApy \cite{Ebrahim2013, Megchelenbrink2014} was used to obtain random points within the feasible space of the model while fixing the fluxes of the reactions for which we had estimations from experimental data. 
\section*{Acknowledgments}
ADM acknowledges support from the European REA, Marie Skłodowska-Curie Actions, grant agreement no. 101131463 (SIMBAD).
D.D.M. acknowledges financial support from the grants
PIBA$\_$2024$\_$1$\_$0016 (Basque Government) and Project PID2023-146408NB-
I00 funded by MICIU/AEI/10.13039/501100011033 and by FEDER, UE. A.F.F. acknowledges support from the Predoctoral Training
Program for Non-Doctoral Research Personnel of the Basque Government's Department of Education

\bibliography{references}

\clearpage

\renewcommand{\thefigure}{S.\arabic{figure}}    
\setcounter{figure}{0}

\renewcommand{\thetable}{S.\arabic{table}}    
\setcounter{equation}{0}



\begin{widetext}
    
\centerline{{\bf SUPPORTING INFORMATION}}





\section{From isotopic ratios to uptake rates}

To translate isotopic ratios into uptake rates, we assume an exponential balanced growth model (Fig. \ref{FigS1}), meaning that intracellular concentrations relax exponentially (with uptake rate $r$) from an initial value $C(0)=C_0$ to a final value $C(\infty)=C_1$, i.e.
\begin{equation}  
C(t) = C_0 + (C_1 - C_0)(1-e^{-r t})~~.
\end{equation}

Isotopic ratios are defined as the concentration of the heavy ($H$) isotope over the concentration of the light ($L$) isotope, such that the natural (initial) ratio is given by $x_0=C_0^{H}/C_0^{L}$, whereas the labeled (final achievable given the label strength) ratio is $x_1 = C_1^{H}/C_1^{L}$.

\begin{figure}[h]
\begin{centering}
\includegraphics[width=0.5\textwidth]{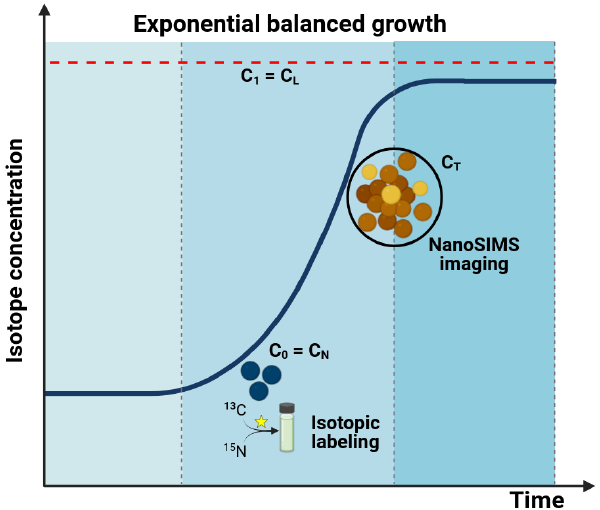}
\caption{Graphical representation of exponential balanced growth with the concentration of labeled isotopes (e.g. \textsuperscript{13}C or \textsuperscript{15}N) growing exponentially for each individual cell. The initial concentration $C_0$ corresponds to the natural isotopic concentration ($C_N$) and it can evolve exponentially to a maximum $C_1$ that cannot exceed the label concentration $C_L$. Cells were sampled for nanoSIMS imaging at a time $T$ before saturation, when the isotopic concentration is $C_T$. \label{FigS1}} \end{centering}
\end{figure}

To obtain an equation for the dynamics of ratios, we start from the fact that, at a generic time $t$,
\begin{equation}  \label{kjhgfhji}
x(t)\equiv\frac{C^{H}(t)}{C^{L}(t)} = \frac{C_0^{H} + ( C_1^{H} - C_0^{H}) (1-e^{-rt})} {C_0^{L} + ( C_1^{L} - C_0^{L}) (1-e^{-rt})}=\frac{x_0 + \left( \frac{C_1^{H}}{C_0^{L}} - x_0\right) (1-e^{-rt})} {1 + \left( \frac{C_1^{L}}{C_0^{L}}- 1\right) (1-e^{-rt})} ~~.
\end{equation}
Note that $x\to x_0$ (resp. $x\to x_1$) for $t\to 0$ (resp. $t\to\infty$), as it should. The superscripts $H$ and $L$ denote the heavy and light isotopes, respectively. If the sum of the concentration of the two isotopes remains constant (as reasonable in view of mass balance), then
\begin{equation}
C_0^{H}+C_0^{L} = C_1^{H} + C_1^{L} ~~,
\end{equation}
implying
\begin{equation}\label{kjhgfdghjk}
\frac{C_1^{H}}{C_0^{L}}+\frac{C_1^{L}}{C_0^{L}}  = 1+x_0~~.
\end{equation}
In addition, we have 
\begin{equation}\label{ytfdtyugjhm}
\frac{C_1^{H}}{C_0^{L}}=x_1\,\frac{C_1^{L}}{C_0^{L}}~~.
\end{equation}
Putting (\ref{kjhgfdghjk}) and (\ref{ytfdtyugjhm}) together we find
\begin{equation}\label{uytdgfhj}
\frac{C_1^{L}}{C_0^{L}}=\frac{1+x_0}{1+x_1}~~~\text{and}~~~ \frac{C_1^{H}}{C_0^{L}}  = \frac{x_1(1+x_0)}{1+x_1}~~.
\end{equation}
Plugging (\ref{uytdgfhj}) into (\ref{kjhgfhji}) we finally obtain
\begin{equation}\label{jhgfdxgh}
x(t) = \frac{x_1 (1+x_0)+(x_0-x_1) e^{-rt}} {1 + x_0 + (x_1-x_0) e^{-rt}}  ~~,
\end{equation}
which describes the time evolution of $x(t)$ from $x(0)=x_0$ to $x(\infty)=x_1$. 

It follows that, if the isotopic ratio measured at time $T$ is $x(T)=x$, the uptake rate $r$ is  given by
\begin{equation}
r =  -\frac{1}{T} \log{\frac{(x_1-x)(1+x_0)}{(x_1-x_0)(1+x)}} ~~.
\end{equation}
\begin{figure}[h]
\begin{centering}
\includegraphics[width=0.55\textwidth]{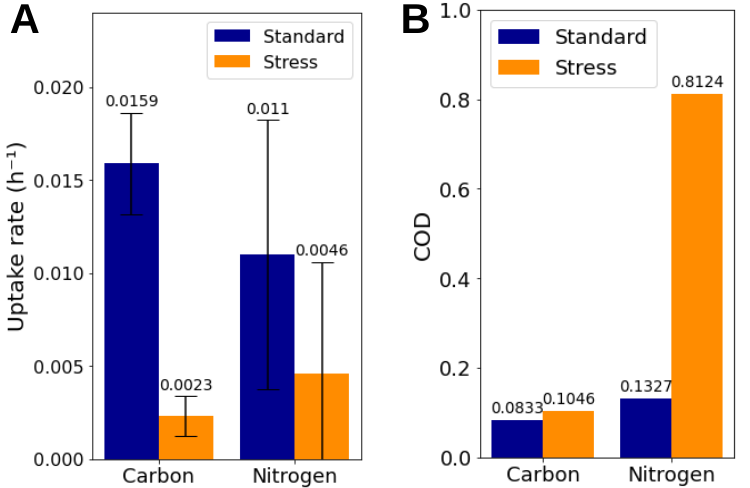}
\caption{\textbf{Statistics of the carbon and nitrogen nanoSIMS-derived uptake rates shown in Figure 1 panel C. } 
\textbf{A.} Average and standard deviation for carbon and nitrogen uptake rates in standard and stress conditions. 
\textbf{E.} Coefficient of Deviation (COD) of the data in A.  \label{FigS2}}  
\end{centering}
\end{figure}

This is the formula reported in the inset of Fig. 1B of the Main Text. Fig \ref{FigS2} shows the mean uptake rates of the two populations along with an estimate of their heterogeneity.

\section{From uptake rates to maximal single-cell growth rates}

In most nanoSIMS isotopic labeling studies, the carbon uptake rate is used as the ``C-based speciﬁc division rate'' \cite{Berthelot2018}. Here, we attempt a more precise quantification of the single-cell growth rate by making a distinction between the uptake rate of an element (e.g., carbon) and the actual growth rate of the cell. 

The starting observation is that, for each cell, the uptake rate $r$ of an element, which coincides with the ratio between that element's uptake flux ($u$) and its concentration ($c$), must equal the rate at which the compound is incorporated into the biomass (i.e., the growth rate $\lambda$) plus a turnover rate ($\gamma$) due to atoms meeting a different fate (including flowing out of the cell). Moreover, under balanced growth, $\lambda$ should be element independent. As a consequence, for carbon and nitrogen one should have
\begin{gather}
r_C \equiv u_C / c_C  = \lambda + \gamma_C ~~,\\
r_N \equiv u_N / c_N = \lambda + \gamma_N~~.
\end{gather}
 
In other terms, uptake rates, in principle, have a common part associated with the growth rate, while they differ by the turnover.

Note that (reasonably) the cell-to-cell variability in $C$ and $N$ levels is much smaller than the cell-to-cell variability of their respective uptake rates, since the latter encodes for the variability from enzyme and transporter levels. Unique values of carbon and nitrogen concentrations were therefore taken from experimental measurements of cell composition in each of the conditions, i.e. Standard conditions: 41.07 mmol of C/g$_{DW}$ and 8.22 mmol of N/g$_{DW}$, stress conditions: 42.01 mmol of C/g$_{DW}$ and 6.32 mmol of N/g$_{DW}$. C/N ratios  measured with Fourier transform infrared spectroscopy, FTIR, using ammonium formate solution. This allows to directly connect uptake rates to uptake fluxes in single cells.

In addition, because turnover rates are non-negative, the above equations imply that
\begin{gather}
 \lambda \leq r_C   ~~,\\
 \lambda \leq r_N ~~.
\end{gather}
As both inequalities have to be satisfied, $\lambda$ cannot exceed the minimum between $r_C$ and $r_N$. That is, given the carbon ($r_{C,i}$) and nitrogen ($r_{N,i}$) uptake rates of cell $i$, its growth rate $\lambda_i$ must be such that
\begin{equation}\label{ugdfhjk}
0   \leq \lambda_i \leq \lambda_{\max,i}\equiv \min \{r_{C,i},r_{N,i}\} ~~. 
\end{equation}
In this way, one obtains an upper bound for the growth rate of each cell within the population.

\section{Bulk growth rate}

The condition (\ref{ugdfhjk}) can be used to obtain a refined estimate of the population (bulk) growth rate. 

A first estimate of the bulk growth rate of the Synechocystis population was obtained from experiments by combining cell counting with fluorescence microscopy and optical density measurements at a wavelength of 750 nm. This returned the values $\mu\pm\sigma=0.011\pm 0.001$ h$^{-1}$ and $\mu\pm\sigma=0.0018\pm 0.0004$ h$^{-1}$ for the standard and stressed conditions, respectively. 

To refine these values, we can account for (\ref{ugdfhjk}) assuming that, in each condition, single-cell growth rates follow approximately a Gaussian distribution with mean $\mu$ and standard deviation $\sigma$, with truncations dictated by (\ref{ugdfhjk}). Corrected bulk growth rates can finally be computed as
\begin{equation}
\lambda_{exp} = \frac{\int_{0}^{\lambda_{\max}}\frac{\lambda e^{-\frac{(\lambda-\mu)^{2}}{2\sigma^{2}}}}{\sqrt{2\pi\sigma^{2}}}d\lambda}{\int_{0}^{\lambda_{\max}}\frac{ e^{-\frac{(\lambda-\mu)^{2}}{2\sigma^{2}}}}{\sqrt{2\pi\sigma^{2}}}d\lambda}~~.
\end{equation}
We obtained $0.009$ h$^{-1}$ and $0.0013$ h$^{-1}$ for the standard and stressed conditions, respectively.

\section{Maximum-entropy estimation of single-cell growth rates}

In principle, any set of single-cell growth rates satisfying the bounds (\ref{ugdfhjk}) is acceptable. For a better estimate we however resort to the Maximum Entropy principle \cite{Jaynes2003}, according to which the most unbiased estimate for the distribution of a random variable is obtained by maximizing the entropy functional subject to constraints encoding for available knowledge. In our case, we can inform entropy maximization by imposing that the average of single-cell growth rates matches the corrected estimate $\lambda_{exp}$, that is 
\begin{equation}\label{kjhgfdhnmj}
\frac{1}{N}\sum_{i=1}^N \langle  \lambda_i \rangle = \lambda_{exp}  ~~, 
\end{equation}
with $N$ the number of cells. Maximization of the entropy functional $H[P]=-\int P(\lambda)\ln P(\lambda)d\lambda$ subject to a constraint on the average growth rate leads to the single-cell growth-rate distribution \cite{de2018introduction}
\begin{equation}\label{kjhfgdvbgn}
P(\lambda_i ) = \frac{e^{\beta \lambda_i}}{Z_i(\beta)}\quad~~ \quad (0   \leq \lambda_i \leq \lambda_{\max,i})~~,
\end{equation}
known in physics as Boltzmann distribution. The function
\begin{equation}
Z_i(\beta)\equiv\int_0^{\lambda_{\max,i}}e^{\beta\lambda_i}d\lambda_i=\frac{1}{\beta}\left(e^{\beta\lambda_{\max,i}}-1\right)
\end{equation}
ensures normalization, while the factor $\beta$ represents the Lagrange multiplier enforcing the constraint, to be determined by solving Eq. (\ref{kjhgfdhnmj}) for $\beta$. More precisely, because
\begin{equation}
\langle\lambda_i\rangle\equiv\int_0^{\lambda_{\max,i}}\lambda_i\,e^{\beta\lambda_i}d\lambda_i=\frac{\partial}{\partial\beta}\ln Z_i(\beta)=\frac{\lambda_{\max,i}}{1-e^{-\beta\lambda_{\max,i}}}-\frac{1}{\beta}~~,
\end{equation}
equation (\ref{kjhgfdhnmj}) takes the form
\begin{equation}
\frac{1}{N}\sum_{i=1}^N  \frac{\lambda_{\max,i }}{1- e^{-\beta \lambda_{\max,i} } } =  \lambda_{exp} +\frac{1}{\beta}~~,
\end{equation}
which can be solved numerically for $\beta$ for any given $N$, $\lambda_{exp}$ and $\{\lambda_{\max,i}\}_{i=1}^N$ (see Fig \ref{FigS3}). 

\begin{figure}[h]
\begin{centering}
\includegraphics[width=1\textwidth]{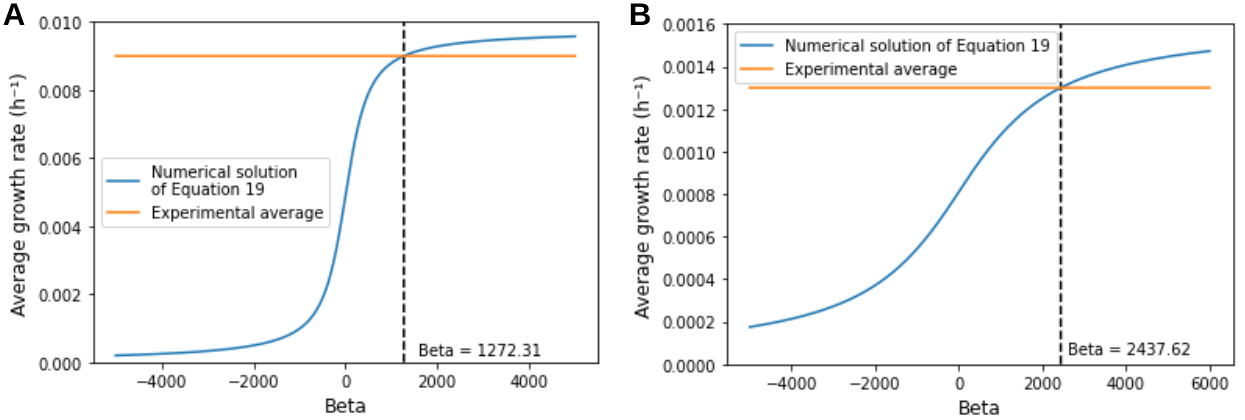}
\caption{Numerical solution for beta in standard (A) and stress (B) conditions } \label{FigS3}
\end{centering}
\end{figure}

Once the solution $\beta^\star$ is found, the growth rate of cell $i$ can be estimated to be given by the mean of (\ref{kjhfgdvbgn}), that is 
\begin{equation}\label{jhgfdghjj}
\langle \lambda_i \rangle = \frac{\beta^\star}{e^{\beta^\star\lambda_{\max,i}}-1}\int_{0}^{\lambda_{\max,i}}\lambda_i \,e^{\beta^\star \lambda_i} d\lambda_i~~.
\end{equation}
Notice that the value thus found depends on both $\lambda_{\max,i}$ (which is cell-dependent) and $\beta^\star$ (which is instead a population-level variable, as it enforces the global constraint (\ref{kjhgfdhnmj})). Likewise, one can estimate the variance $\sigma_i^2\equiv\langle \lambda_i^2\rangle-\langle \lambda_i\rangle^2$ from (\ref{jhgfdghjj}) and
\begin{equation}
\langle \lambda_i^2\rangle = \frac{\beta^\star}{e^{\beta^\star\lambda_{\max,i}}-1}\int_{0}^{\lambda_{\max,i}}\lambda_i^2\,e^{\beta^\star \lambda_i} d\lambda_i~~.
\end{equation}
This assumes for sake of simplicity that growth rates are independent. In a different framework, akin to the microcanonical ensemble in statistical mechanics, we could assume that the population average is strictly equal to the measured bulk growth rate
\begin{equation}
\frac{1}{N}\sum_{i=1}^N  \lambda_i  = \lambda_{exp}  
\end{equation}
In this case we have no simple analytical formula and the space shall be sampled numerically (even if its formal analogy with the simplex case would suggest analytical feasibility). We checked with a Markov chain monte carlo  method that this framework gives very similar results with respect to the maximum entropy above for average single cell growth rates. The maximum entropy framework shall be anyway considered more conservative and better justified statistically.

\section{Genome-scale metabolic modeling}

The exhaustively curated genome-scale metabolic model of Synechocytis sp. PCC6803 (iJN678) published by \cite{Nogales2012} served as the basis for the metabolic modeling. Some minor modifications were made for the purposes of this work. A different model was used for each of the conditions in order to account for the different C-N ratios observed experimentally. To achieve these  values the mass balance of the biomass equation was slightly modified by adjusting the stoichiometric coefficients. In addition, a diffusion equation for the releasing of the excess of nitrogen was included in the model to allow for the simulation of the turnover of this element. 
The bounds for the exchange reactions were calculated from media composition and considering diffusion constraints and crowding effects. Diffusion theory holds that the uptake flux of a cell is limited by 
\begin{equation}
u \leq \frac{4\pi cDR}{m}   
\end{equation}
where $c$ is the concentration (mmol/L), $D$ is the diffusion constant ($\mu m^2$/s), $R$ is the linear size of the cell ($\mu m$) and $m$ is the cell mass (g).
If crowding effects are considered, for many cells: 
\begin{equation}
u\rho \int_{R}^{L}\frac{dV}{R} = u\rho \int_{R}^{L}4\pi R dR \leq \frac{4\pi cD}{m} 
\end{equation}
where $\rho = \frac{N}{4/3\pi L^3}$ is the cell density ($N$ = number of cells) and $L$ is the length of the system that is approximately 2cm.
\begin{equation}
u \leq \frac{2 cDR}{m\rho(L^2-R^2)} = \frac{\frac{8}{3}\pi cD \frac{L^3}{L^2-R^2}}{Nm}   
\end{equation}
The length of the system is orders of magnitude bigger than the radius of the cells ($L>>R$) which derives in the following expression for the maximum uptakes: 
\begin{equation}\label{eq:19}
u \leq \frac{\frac{8}{3}\pi cDL}{Nm}   
\end{equation}
Given Eq \ref{eq:19}, the bounds in table below were calculated for each compound in the medium. 

The maximum photon uptake was calculated from the experimental irradiance of 50$\mu$E taking mass and diameter of an average Synechocystis cell to be 150fg (\cite{QIAN2017276}) and 2.1 $\mu$m (\cite{10.1371/journal.pone.0189130}) respectively , and assuming the maximal efficiency of photosynthesis to be 6 \% (\cite{ZHU2008153}). Additionally, it was multiplied by a factor of 0.6 to correct forelight-dark cycles and a factor 0.7 to account for light attenuation according to the Beer-Lambert law.

Additionally, an ATP maintenance reaction was included in the model to account for the non-growth-associated maintenance (NGAM) ATP demand. The minimum flux bound was estimated based on protein turnover, the primary maintenance cost. We assumed that a typical cyanobacterium replaces half of its protein content every 12 hours. A single cell of this type contains approximately $10^6$ proteins, each with an average length of 200 amino acids. The energy cost for protein synthesis includes 1 ATP for initiation and 3 ATP per peptide bond. Using these values, we estimated the ATP required for protein turnover $(\gamma_{p}$) as: 
\small{\begin{equation}\label{eq:20}
 \gamma_{p} = \frac{10^6}{2} proteins \times \left( 1 ATP + 3\frac{ATP}{pb} \times 199\frac{pb}{protein} \right ) \times\frac{1mmol}{6\times 10^{26} ATP}\times\frac{1}{3\times10^{-12}g/cell}\times\frac{1}{12h} 
\end{equation}}
This calculation results in a lower bound for the ATP maintenance reaction of 
\begin{equation}
u_{ATPM}\geq  0.13 mmol ATP /gDWh.
\end{equation}

Flux balance analysis (FBA) \cite{Orth2010} was applied to calculate the maximum growth rate. In both cases $\lambda_{max}\sim0.018$h$^{-1}$ indicating that the limiting factor is light. To calculate the feasible space, the growth rate was fixed at different values iterating between 0 and the maximum growth rate (which is mainly determined by the upper bound of photon flux) and the objective function was set to be either the carbon or the nitrogen uptake. At each value of the growth rate, FBA was employed to calculate the maximum and minimum values for the uptakes.

\begin{table}[h]
\centering
\caption{Flux bounds in $mmol$·$gDW^{-1}$·$h^{-1}$}
\begin{tabular}{lrrr}
uptake/medium & standard & stressed \\
\hline
$CO_2$ & $1.3$ & $664$ \\
$NH_4$ & $210$ & $605$ \\
$SO_4$ & $1.96$ & $654$ \\
$O_2$ & $1000$ & $0$ \\
$Photons$ & $9$ & $9$ \\
\hline
\end{tabular}
\end{table}

\section{Clustering analysis}

A hierarchical clustering analysis \cite{murtagh2012algorithms} was conducted to identify subpopulations with distinct metabolic profiles under the two conditions. As a first step, the K-means algorithm \cite{MacQueen1967} was applied using the logarithm of the turnover rates as the key parameters. This is given by 
\begin{equation}
\log\gamma = \log{(u - {\lambda}c)}+\mathrm{constant}~~.
\end{equation}
This initial clustering yielded two groups that differed in their optimization criteria, prioritizing either carbon or nitrogen utilization. Subsequently, each cluster was  subdivided using the same algorithm, this time based on growth rates.  In other words,  first cells were clustered together according to their turnover rates (Step 1, 2d), then within each cluster  cells were clustered again according to their growth rate (Step 2, 1d). Step 1 gives the clusters  red+purple and green+cyan; Step 2 separates red from purple and green from cyan.

\section{Excessive acid exudation from the isolated cell model leads to unviable pH}
The metabolic flux analysis performed by fixing the growth rate and C-N uptakes of single cells in the genome scale model returns exceedingly high values for acid cell exudation (see Fig \ref{FigS4}, where the average exudation amounts at $u_e=0.06$mmol/gh,  that is $15\%$ of the amount fixed into biomass $0.42$mmol/gh), since in considering isolated cells, the carbon uptake shall be fixed to the one of $CO_2$. 
Upon considering a cell density $\rho=1 g/l$ and a waiting time of $T=21h$, the concentration of acids (in carbon monomers) will be $c_A = \rho T  u_e \sim 1.3$mM.
This concentration is beyond the buffering capacity of the medium (BG11, that has a phosphate and carbonate buffers with concentrations of $c_p=0.23$mM and $c_b=0.19$mM respectively).
\begin{figure}[h]
\begin{centering}
\includegraphics[width=1\textwidth]{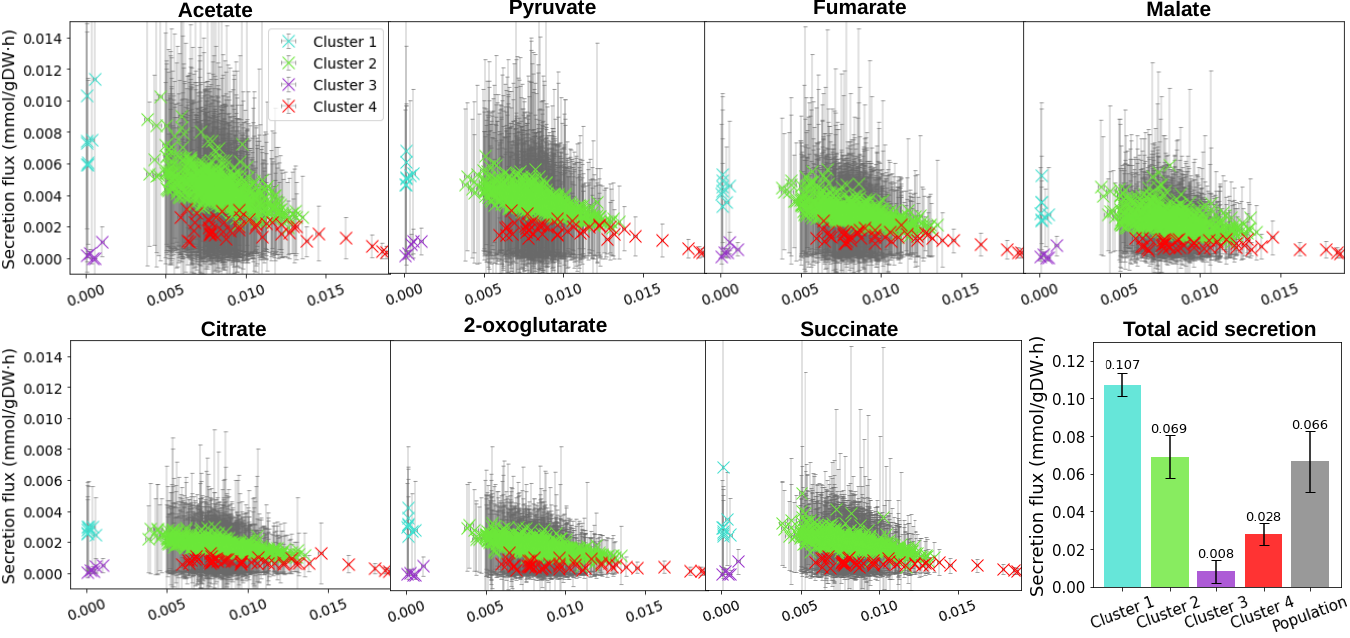}
\caption{Scatter plots showing the secretion fluxes of various acids included in the model (before considering exchanges, ie isolated cells), with colors representing different clusters. The bar plot in the bottom right illustrates the average total secretion flux for each acid, both within individual clusters and across the entire population. } \label{FigS4} 
\end{centering}
\end{figure}

We performed titration computations that we report here in Fig \ref{FigS5}, where the pH as a function of the level of exudation is shown, highlighting with dots the levels corresponding to the model used for inference taking into account the exchanges or not. In the latter case (isolated single cells) the pH drops to unviable values $pH<5$.  

The computations were performed by considering a buffer made of phosphate and bicarbonate at equilibrium with acetic acid (approximated to be the only one), see also (\cite{beard2008chemical}). 
We have the equations for the chemical equilibria
\begin{eqnarray}
[H^+][OH^-] = k_w  \quad  \frac{[H^+][A^-]}{[HA]} = k_l \\
\frac{[H^+][HCO_3^-]}{[H_2 CO_3]} = k_1  \quad  \frac{[H^+][CO_3^{2-}]}{[HCO_3^-]} = k_{12} \\ 
\frac{[H^+][H_2PO_4^-]}{[H_3 PO_4]} = k_{1b}  \quad  \frac{[H^+][HPO_4^{2-}]}{[H_2PO_4^-]} = k_{12b}  \quad \frac{[H^+][PO_4^{3-}]}{[HPO_4^{2-}]} = k_{123b} 
\end{eqnarray}
then we have the mass conservation equations (where we approximately consider the hydroxides of sodium and potassium completely dissociated) 
\begin{eqnarray}
\left[A^-\right] + \left[HA\right] = c_A \\
\left[H_2 CO_3\right] + [HCO_3^-] +  [CO_3^{2-}] = c_b \\
\left[H_3 PO_4\right] + [H_2PO_4^-] +  [HPO_4^{2-}] + [PO_4^{3-}]  = c_p \\
\left[Na^+\right] \sim 2 c_b \quad [K^+] \sim 2 c_p
\end{eqnarray}
and finally the equation for the charge balance
\begin{equation}
[H^+]  + [Na^+] + [K^+] = [OH^-] +   [HCO_3^-] + 2  [CO_3^{2-}] + [H_2PO_4^-]  + 2[HPO_4^{2-}]+ 3[PO_4^{3-}] 
\end{equation}
The last equation leads to the following implicit equation between hydrogen ion concentration $x=[H^+]$ and acid concentration $y=c_A$
\begin{equation}
\begin{split}
y = \left(1. + \frac{x}{k_{l}}\right) \cdot \bigg[ &x - \frac{k_{w}}{x} \\
& - \frac{c_b \left(\frac{k_{1}}{x} + \frac{2 \cdot k_{12}}{x^2}\right)}
          {1 + \frac{k_{1}}{x} + \frac{k_{12}}{x^2}} \\
& - \frac{c_{p} \left(\frac{k_{1b}}{x} + \frac{2 \cdot k_{12b}}{x^2} + \frac{3 \cdot k_{123b}}{x^3}\right)}
          {1 + \frac{k_{1b}}{x} + \frac{k_{12b}}{x^2} + \frac{k_{123b}}{x^3}}  \\
& + 2 (c_{p} + c_b) \bigg]
\end{split}
\end{equation}
Parameters, with the dissociation constants for the carbonic phosphoric and acetic acid are reported in the table below (see for instance \cite{atkins2023atkins}). 
\begin{table}[h]
\centering
\caption{Parameter values}
\begin{tabular}{lcc}
\hline
Parameter & Value (units) & description \\  
\hline
$c_p$ & 190 $\mu$M  & phosphate buffer conc.\\
$c_{b}$ & 230 $\mu$M & bicarbonate buffer conc.\\
$k_{w}$ & 0.01 $\mu$M\textsuperscript{2} & water dissoc. const.\\
$k_{l}$ & 17.7 $\mu$M & acetic acid dissoc. const.\\
\hline
$k_{1}$ & 0.45 $\mu$M & carbonic acid dissoc. const. (1) \\
$k_{12}$ & 0.00002115 $\mu$M & carbonic acid dissoc. const. (2) \\
\hline
$k_{1b}$ & 6900 $\mu$M & phosphoric acid dissoc. const. (1) \\
$k_{12b}$ & 428 $\mu$M & phosphoric acid dissoc. const. (1) \\
$k_{123b}$ & 0.000238843 $\mu$M & phosphoric acid dissoc. const. (1) \\
\hline
\end{tabular}
\end{table}
We consider sodium and potassium cations completely dissociated.
We neglect modifications of the medium due to the carbon fixation by the cells, and/or changes induced by the autoclaving. We do point out however that they would lead to a further decrease of the bicarbonate buffer capacity and of the pH, thus confirming the unviability of the exudation levels implied by an isolated single cell model that does not take into account metabolic exchanges.  

\begin{figure}[h]
\begin{centering}
\includegraphics[width=0.6\textwidth]{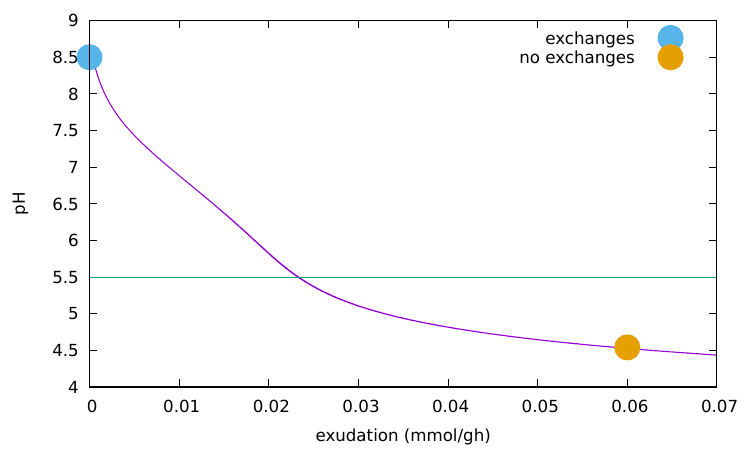}
\caption{pH as a function of average exudation level from titration computations with known values of the buffering medium BG11. Dots are the points } \label{FigS5}
\end{centering}
\end{figure}

\section{Inferring cell-to-cell couplings}

Consider a population of mixotrophic cells of Synechocystis growing in HCO3 as the only carbon source in the medium. Once inorganic carbon enters the cell it is fixed using the energy harvested from light (see Fig. \ref{FigS6}). This fixed carbon can be further split in three different directions. It can be incorporated to the biomass and therefore contributing to increase the growth rate. It can also be released to the environment in this same fixed form. Finally, it can be used in respiration to get additional ATP.  In this way, we can imagine a population of cells that is actively interacting through acid (fixed carbon) exchange: one part of the population acts as carbon fixers and the other presenting a more heterotrophic behavior growing at the expense of the already fixed carbon released by other cells.

The variables of the model are the following five reaction fluxes (per cell): photosynthesis $\gamma$, import of inorganic carbon $u_{co2}$, respiration $u_{o2}$, import/export of organic/fixed carbon $f_a$ and biomass growth rate $\lambda$.

Let also fix some constants for the stoichiometry of reactions: 
\begin{itemize}
    \item $a=6 \xrightarrow{}$ ATP consumed per fixed C
    \item $b=5 \xrightarrow{}$ ATP generated per C in respiration
    \item $\alpha=53 \xrightarrow{}$ ATP consumed per C incorporated to the biomass
    \item $\phi=0.4 \xrightarrow{}$ ATP generated per unit of photon
    \item $\xi=42 mmol/gDW \xrightarrow{}$ biomass content of carbon
\end{itemize}

\begin{figure} [h]
\begin{centering}
\includegraphics[width=0.5\textwidth]{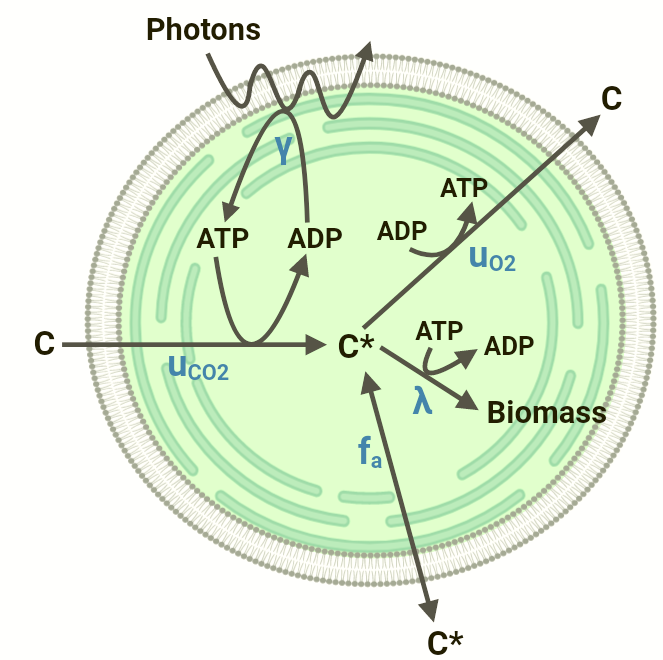} 
\caption{Model scheme of the simplified metabolic network of {\it Synechocystis}. This is composed of five reactions: Photosynthesis $\gamma$, import of inorganic carbon $u_{co2}$, respiration $u_{o2}$, import/export of organic/fixed carbon $f_a$ and biomass growth rate $\lambda$}\label{FigS6}
\end{centering}
\end{figure}
With this scheme, we can define some equations that describe the model. First, we will consider the carbon balance (both $u$ and $f$ denote fluxes and therefore have $mmol$·$gDW^{-1}$·$h^{-1}$ units) :
\begin{gather}
    u_{CO_2} + f_a - u_{O_2} = \xi \lambda ~~,\\
    u_{C_{IN}} = u_{CO_2} + f_a \Theta(f_a>0) ~~,\\
    u_{C_{OUT}} = u_{O_2} - f_a \Theta(f_a<0)~~,
\end{gather}
so that
\begin{equation}
    u_{C_{IN}} - u_{C_{OUT}} = u_{CO_2} - u_{O_2} + f_a = \xi \lambda~~.
\end{equation}
We have independent estimates of the net carbon uptake ($u_{C_{IN}}-u_{C_{OUT}} = \xi \lambda$) for each individual cell. We can write down an equation for the energy balance involving ATP:
\begin{equation}
    f_{ATP} = -a  u_{CO_2} + b u_{O_2} + \phi \gamma - \alpha \lambda
\end{equation}
We also consider that the C fixation and the oxygen flux for respiration must be positive. On the other hand, the flux of acids for a cell living alone must be negative, since there is no source of acids in the medium. Also, the amount of absorbed photons must be between some min/max and the ATP flux should be positive:
\begin{gather}
    0 \leq u_{CO2} \leq u_{CO2, max} ~~,\\
    0 \leq u_{O2} \leq u_{O2, max} ~~,\\
    0 \leq \gamma \leq \gamma_{max} ~~,\\
    f_{ATP} \geq 0 ~~,\\
    f_a \leq 0~~. 
\end{gather}
In the case of many cells, the sum of all the acid fluxes must be negative as well which give us the constraint: 
\begin{equation}
f_{a,i} + \sum_j \frac{f_{a,j}}{d_{ij}/R} \leq 0  \qquad(\forall i)~~.
\end{equation}

For the case of just one cell, the growth rate increases with the photon flux up to a maximum described by FBA that corresponds to the case where there is no acid release ($f_a = 0$) and no respiration ($u_{O2} = 0)$ and the photon flux equals the max ($\gamma=\gamma_{max}$):
\begin{equation}
    \lambda_{max} = \frac{\phi\gamma_{max}}{\alpha+a\xi} = 0.013\,\mathrm{h}^{-1}~~.
\end{equation}

Cells are placed in a liquid medium where the only source of carbon is bicarbonate in equilibrium with CO2 in the atmosphere. Therefore, the concentration of acids in the medium depends only on the cells that release and intake them at different rates, which we assume to remain constant over time. After the system has reached equilibrium, the acid concentration profile should satisfy the Laplace equation $\nabla ^{2}c(\vec{r} ) = 0$, where $c(\vec{r})$ is the concentration of acids at position $\vec{r}$. A solution of this equation for point-like spherical sources and sinks at positions $\vec{r_i}$ is
\begin{equation}
    c(\vec{r}) = c_{\infty} - \frac{1}{4\pi D}\sum_{i=0}^{N} \frac{f_{a,i}}{|\vec{r}-\vec{r_i}|}
\end{equation}
Where N is the number of cells, D is the diffusion coefficient for acetate ($1089\mu m^2/s$) and $c_{\infty}=0$, meaning that the concentration of acids in the background is zero.  

In general for $N$ cells we have $N$ constraints for the uptakes  ($R$ is the cell size and $d_{ij}$ is the distance between cell $i$ and $j$)
\begin{equation}
f_{a,i} + \sum_j \frac{f_{a,j}}{d_{ij}/R} \leq 0  \qquad(\forall i)~~,
\end{equation}
which, together with the constraints for single cells, will define a $3N$-dimensional polytope of feasible metabolic states that can be sampled numerically for instance via Monte Carlo methods. One can define optimal states by solving linear programming problems (flux balance analysis, e.g. by requiring the maximization of $\sum_i \lambda_i$) and/or perform inference and quantitative modeling with maximum entropy methods, i.e.  defining inside the space a Boltzmann measure by which
\begin{equation}
\mathrm{Prob}(\textrm{metabolic state}) \propto e^{\beta\sum_i \lambda_i}   
\end{equation}

\begin{figure}
\begin{centering}
\includegraphics[width=1\textwidth]{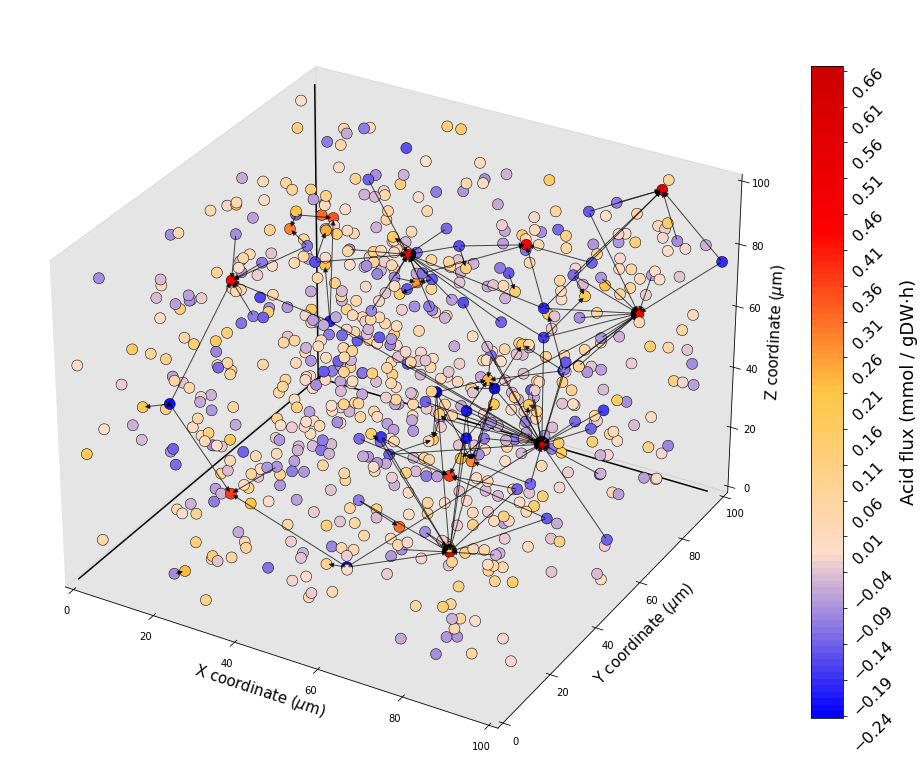}
\caption{\textbf{Interaction network} The axes represent the coordinates of a 3-dimensional space where cells are randomly positioned based on the experimental cell density. Each cell is colored according to the gradient shown on the right, indicating its acid flux—from red (uptake) to blue (secretion). Arrows represent pairwise interactions mediated by acid exchange. \label{FigS7}}  
\end{centering}
\end{figure}
where by the metabolic state we mean the state of the whole population. To simulate our population, we randomly placed cells in a three-dimensional space matching the experimental cell density and considering periodic boundary conditions (see Fig \ref{FigS7} for an example of the population configuration). To exclude the possibility of interactions occurring due to the spatial distribution of cells, we performed the sampling for three different configurations of cell positions and averaged the results over them. 

Once the five fluxes are found for each individual cell, networks of interaction through acid exchange (like the one in Fig \ref{FigS7}) can be generated. For that, we calculate an interaction strength for each pair of nodes (i,j) that have opposite acid flux sign (i: secretor, j:absorber): 
\begin{equation}
S_{i,j} = -\frac{f_{a,i}f_{a,j}}{d_{i,j}}\sum_{k=0}^{k=N}\frac{1}{\frac{f_{k}\theta{(f_k>0)}}{d_{i,k}}}  
\end{equation}
We then establish an interaction between cell i and j when this strength overcomes a threshold that is given by the standard deviation of the acid flux in the population divided by the square root of the number of cells. 
\begin{figure}[h]
\begin{centering}
\includegraphics[width=1\textwidth]{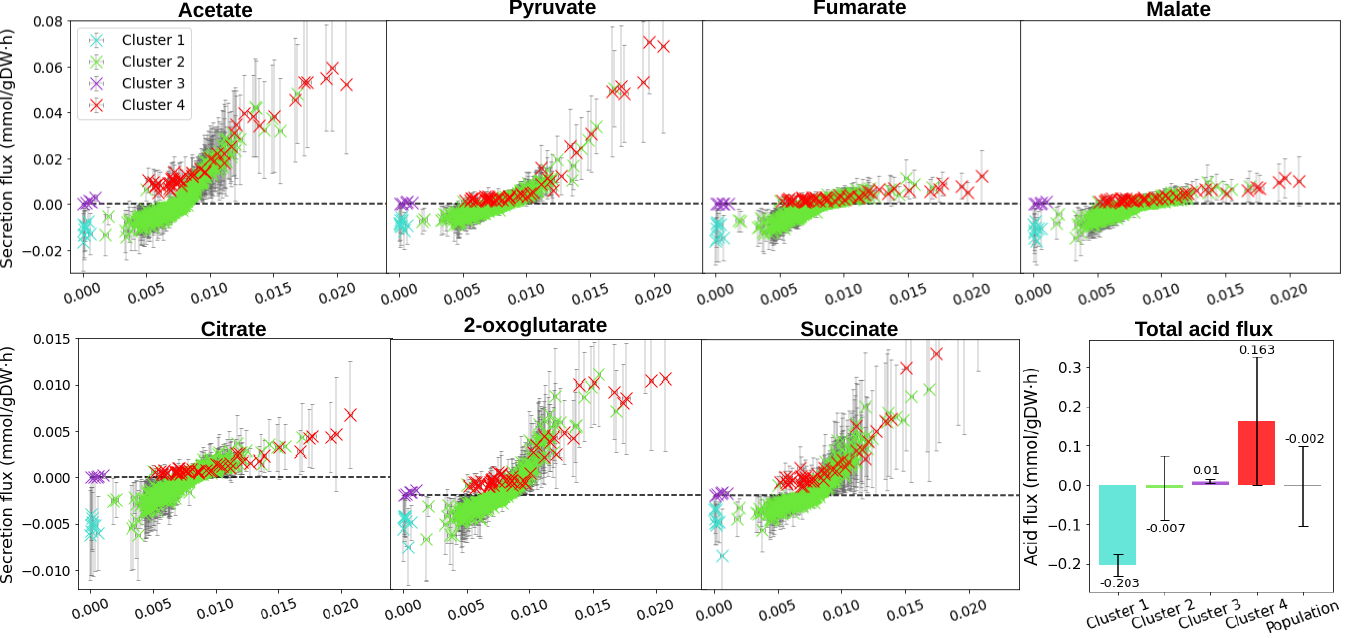}
\caption{Scatter plots showing the fluxes of various acids included in the model after considering exchanges (negative and positive fluxes for secretion and intake respectively), with colors representing different clusters. The bar plot in the bottom right illustrates the average total flux for each acid, both within individual clusters and across the entire population. } 
\label{FigS8}
\end{centering}
\end{figure}

\section{Solution-space sampling algorithm}

In order to obtain random points within the feasible space of the genome-scale model we employed the sampling algorithm implemented in COBRApy \cite{Ebrahim2013, Megchelenbrink2014}. To uniformly sample feasible points inside the multicellular space defined in the section above we employed a modified hit-and-run markov chain to take into account ill-conditioning.
The multi-cellular metabolic model defines a high-dimensional convex polytope ($D = 4N \approx 1000$, where $N$ is the number of cells), and the computational task reduces to characterizing this space through flux distributions weighted by a Boltzmann factor under maximum entropy constraints. This problem belongs to a class of NP-hard challenges---including computing matrix permanents, high-dimensional volumes, or Ising partition functions---that admit polynomial-time numerical solutions via Markov chain Monte Carlo (MCMC) methods~\cite{Simonovits2003HowTC}. For constraint-based metabolic modeling, over-relaxed algorithms like hit-and-run sampling~\cite{Turchin1971} excel when combined with approximate ellipsoidal rounding to address ill-conditioning~\cite{demartino2015}.

Our sampling workflow proceeds as follows:
\begin{enumerate}
\setcounter{enumi}{-1}
\item \textbf{Initialization}: Locate a feasible point $P_0$ inside the polytope (e.g., via relaxation algorithms~\cite{Krauth_1987}).
\item \textbf{Direction sampling}: Given $P_i$, generate a uniformly random unit vector $\hat{n}$ (using Marsaglia's method~\cite{knuth2014art}).
\item \textbf{Boundary intersection}: Solve for $t_1, t_2$ where the line $L(t) = P_i + t \hat{n}$ intersects the polytope boundary.
\item \textbf{Weighted step}: Sample $t^* \in [t_1, t_2]$ from the marginalized Boltzmann distribution along $L(t)$, set $P_{i+1} = P_i + t^* \hat{n}$, and iterate.
\end{enumerate}

\begin{figure}[h]
\begin{centering}
\includegraphics[width=0.8\textwidth]{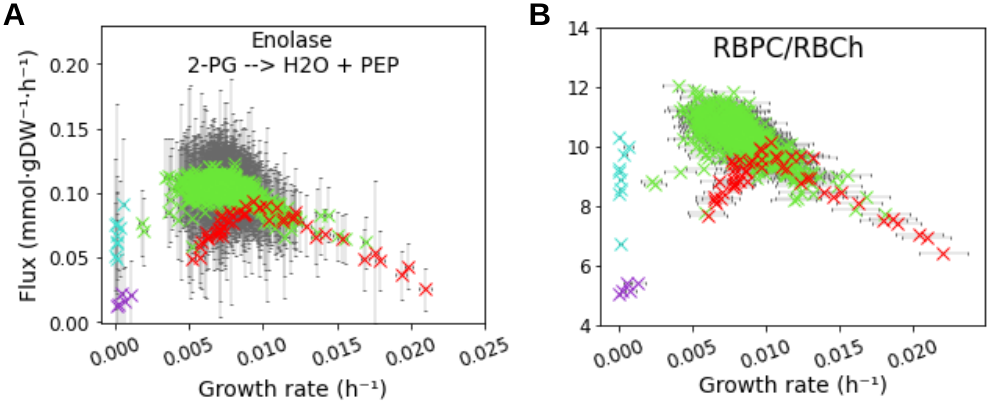}
\caption{\textbf{Results of sampling the genome scale metabolic model informed by the nanoSIMS, maxEnt data and exchanges. } 
\textbf{A.} Scatter plot of the enolase reaction.
\textbf{B.} Rubisco efficiency vs. growth rate  \label{FigS9}}  
\end{centering}
\end{figure}

Notably, ill-conditioning is less severe here than in bulk metabolic networks due to the symmetric structure of the polytope---a product of $N$ identical single-cell flux spaces with diffusion constraints. Code is available in repository ~\texttt{https://github.com/arianferrero/SynechocystisSingleCell.git}

\section{Metabolic pathway enrichment analysis}

To better characterize metabolic differences between clusters, a metabolic pathway enrichment analysis was performed. The activity of each metabolic pathway was quantified for each cluster following a cumulative flux approach. In the iJN678 genome scale metabolic model published by Nogales et al., each reaction is annotated with the metabolic pathway in which it takes place according to databases like KEGG. With this information, it is possible to map every reaction to the metabolic pathway to which it is associated. This allows us to estimate the activity of each metabolic pathway from the results of the sampling algorithm, as the sum of the absolute value of the flux through every reaction associated to it such that, for the example of the tricarboxylic acid cycle (TCA): 
\begin{equation}
     ACT_{TCA} = \sum_i|f_i|  \text{ if } f_i \in TCA
\end{equation}
These results feed the clustermap in Figure 5I of the Main Text. 

\clearpage

\begin{figure}[h]
\begin{centering}
\includegraphics[width=1\textwidth]{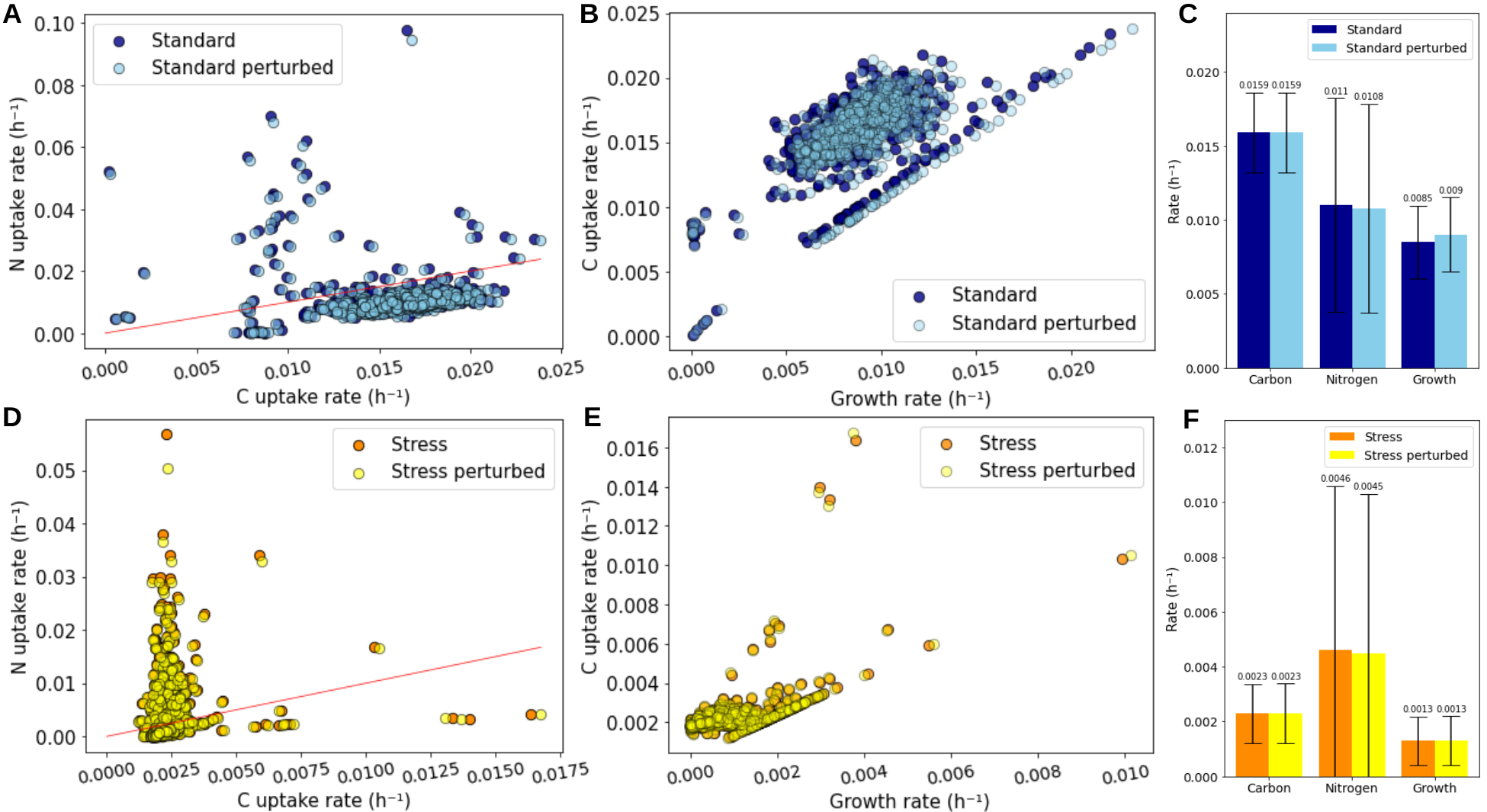}
\caption{\textbf{Sensitivity analysis of SIMS measurement error.} Isotopic ratio data were perturbed by $\pm1\%$ of the original SIMS values to assess robustness.
\textbf{A.} Scatter plot comparing results on elemental uptake rates before and after perturbation under standard conditions. 
\textbf{B.} Scatter plot comparing results on growth rate calculated with maximum entropy before and after perturbation under standard conditions. 
\textbf{C.} Bar plot showing changes in population mean and standard deviation after perturbation under standard conditions.
\textbf{D--F.} Corresponding plots for stress conditions, equivalent to panels A, B and C.
 \label{FigS10}}  
\end{centering}
\end{figure}

\begin{figure}[h]
\begin{centering}
\includegraphics[width=1\textwidth]{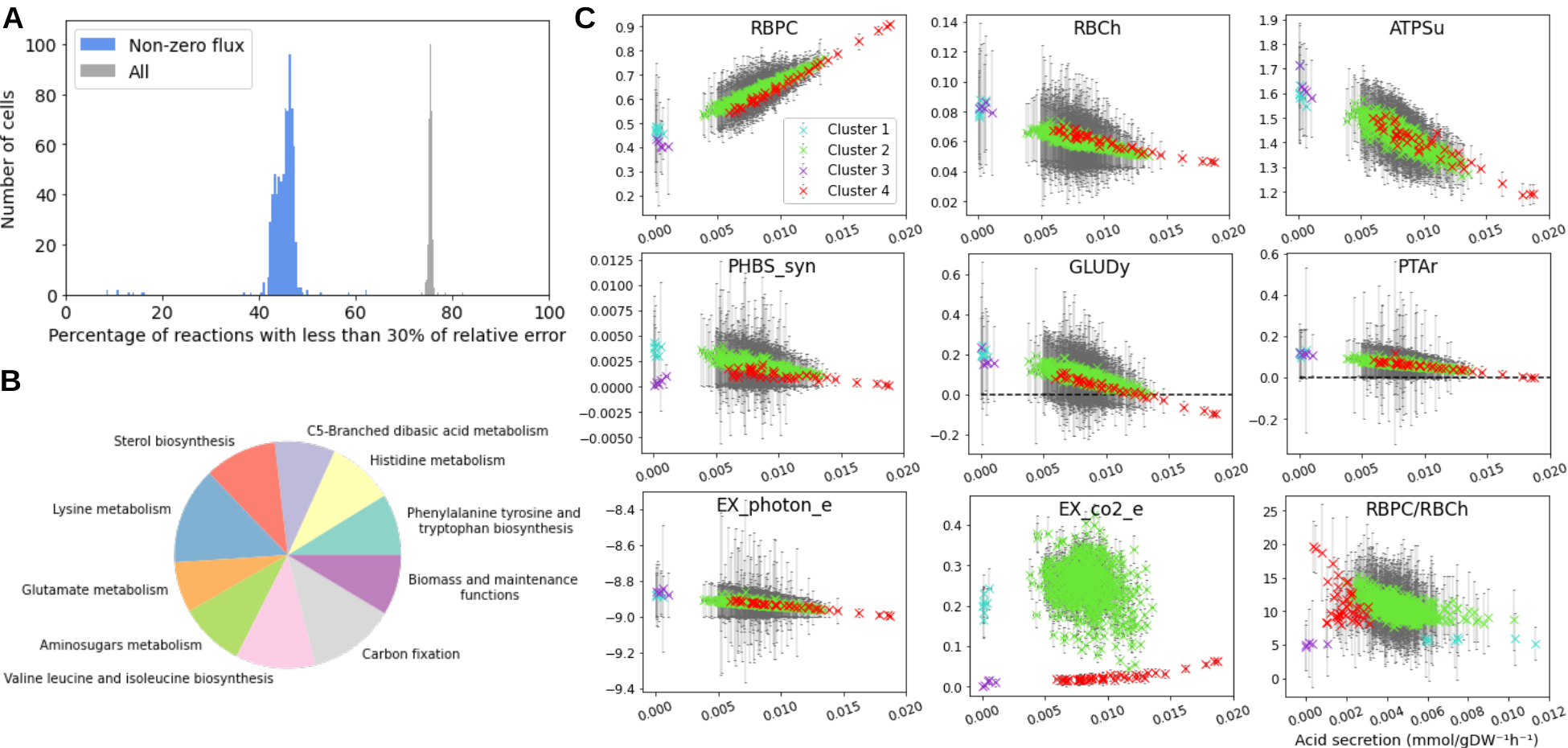}
\caption{\textbf{Results of sampling the genome scale metabolic model informed by the nanoSIMS and maxEnt data, previous to consider exchanges. } 
\textbf{A.} Histogram of the proportion of reactions with less than a 30\% of relative error for each cell. The reactions were consider to be zero if the flux was lower than 0.1\% of the carbon uptake flux. 
\textbf{B.} Piechart showing the metabolic functions that are better resolved in the population. \textbf{C.} Scatter plots of some relevant individual reactions that were chosen for the results of Figure 5 in the Main Text. \label{FigS11}}  
\end{centering}
\end{figure}

\begin{figure}
\begin{centering}
\includegraphics[width=1\textwidth]{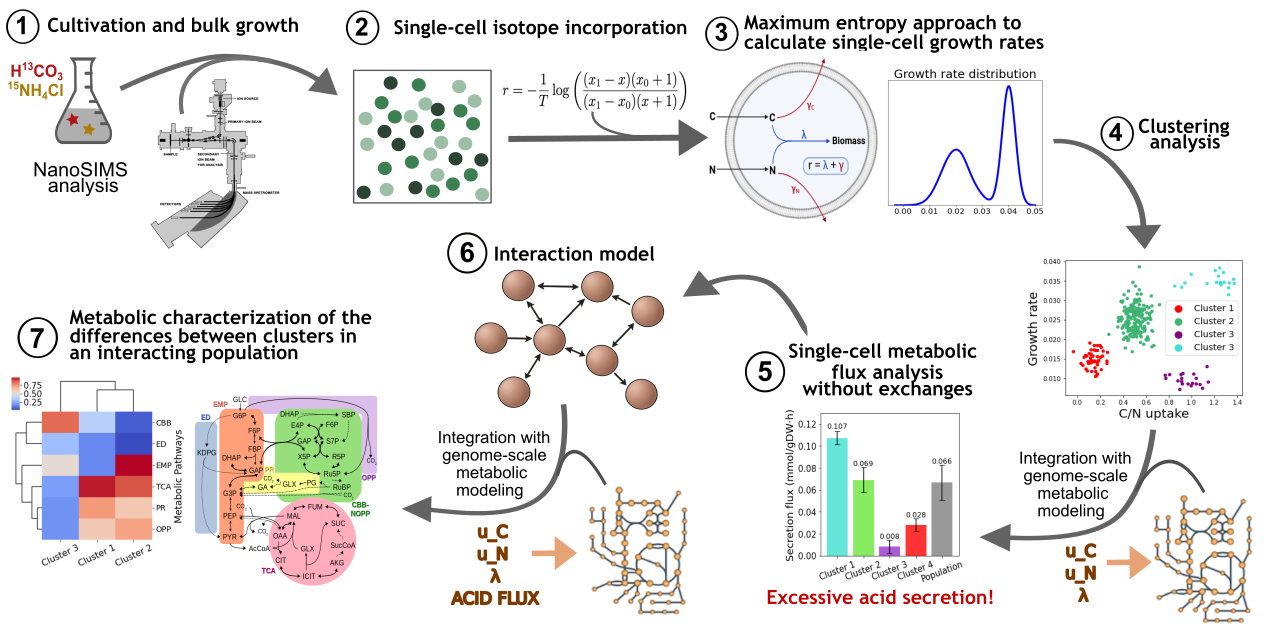}
\caption{\textbf{Workflow}  1) Cells are cultivated and grown 2) in labeled nutrient media with multisotopes  whose accumulation has been measured in single cells with nanoscale mass spectrometry. 3) Single-cell growth rates are inferred from measured uptakes and the bulk growth rate from Maxent and Liebig law. 4) single cell metabolic heterogeneity is evaluated via clustering. 5) results on carbon and nitrogen uptake and growth rates are integrated with constraint-based network modeling to estimate single-cell metabolic fluxes. 6) A simplified model was developed to account for interactions through organic acid exchanges to solve inconsistencies in acid production and feasibility. 7) results of step 2 and 3 were integrated together with the acid fluxes derived from step 6 to solve the single cell metabolism of an interacting population. }
\end{centering}
\end{figure}

\end{widetext}

\end{document}